\documentclass[a4paper]{article}

\usepackage{graphicx}
\usepackage{epstopdf}
\usepackage{epsfig}

\usepackage{amsmath}
\usepackage{graphics,graphicx}
\usepackage{mathrsfs}
\usepackage{color}

\usepackage[margin=1in]{geometry}

\usepackage[T1]{fontenc}
\newcommand{\changefont}[3]{
\fontfamily{#1} \fontseries{#2} \fontshape{#3} \selectfont}

\changefont{ptm}{m}{n}

\usepackage{setspace} 
 \usepackage{graphicx}    

\newcommand \be{\begin{equation}}
\newcommand \ee{\end{equation}}
\newcommand \ba{\begin{eqnarray}}
\newcommand \ea{\end{eqnarray}}

\def\bit{\begin{itemize}}
\def\eit{\end{itemize}}

\usepackage{amsfonts}
\usepackage{amssymb}
\usepackage{graphics}
\usepackage{mathrsfs}
\usepackage{color}

\newtheorem{theorem}{Theorem}[section]

\long\def\symbolfootnote[#1]#2{\begingroup%
\def\thefootnote{\fnsymbol{footnote}}\footnote[#1]{#2}\endgroup} 

\begin{document}

%

\begin{center}
\Large \textbf{Extension of Lorenz Unpredictability}
\end{center}

\vspace{-0.3cm}
\begin{center}
\normalsize \textbf{Marat Akhmet$^{a,} \symbolfootnote[1]{Corresponding Author Tel.: +90 312 210 5355,  Fax: +90 312 210 2972, E-mail: marat@metu.edu.tr}$, Mehmet Onur Fen$^b$} \\
\vspace{0.2cm}
\textit{\textbf{\footnotesize$^a$Department of Mathematics, Middle East Technical University, 06800, Ankara, Turkey}} \\
\textit{\textbf{\footnotesize$^b$Neuroscience Institute, Georgia State University, Atlanta, Georgia 30303, USA}}
\vspace{0.1cm}
\end{center}

\vspace{0.3cm}

\begin{center}
\textbf{Abstract}
\end{center}

\noindent\ignorespaces

It is found that Lorenz systems can be unidirectionally coupled such that the chaos expands from the drive system. This is true if the response system is not chaotic, but admits a global attractor, an equilibrium or a cycle. The extension of sensitivity and period-doubling cascade are theoretically proved, and the appearance of cyclic chaos as well as intermittency in interconnected Lorenz systems are demonstrated. A possible connection of our results with the global weather unpredictability is provided.

\vspace{0.2cm}
 
\noindent\ignorespaces \textbf{Keywords:} Lorenz system; Chaos extension; Sensitivity; Period-doubling cascade; Intermittency; Cyclic chaos

\section{Introduction} \label{butterfly_intro}

In his famous study, to investigate the dynamics of the atmosphere, Lorenz \cite{Lorenz63} built a mathematical model consisting of a system of three differential equations in the following form, 
\begin{eqnarray}
\begin{array}{l} \label{lorenz_system}
\displaystyle \frac{dx_1}{dt}  = -\sigma x_1 + \sigma x_2 \\
\displaystyle \frac{dx_2}{dt}  =  -x_1x_3 +rx_1 -x_2\\
\displaystyle \frac{dx_3}{dt}  = x_1x_2-bx_3,
\end{array}
\end{eqnarray}
where $\sigma, r$ and $b$ are constants.

System (\ref{lorenz_system}) is a simplification of a model, derived by Saltzman \cite{Saltzman62}, to study finite amplitude convection. The studies of Saltzman originate from the Rayleigh-B\'{e}nard convection, which describes heat flow through a fluid, like air or water. In this modelling, one considers a fluid between two horizontal plates where the gravity is assumed to be in the downward direction and the temperature of the lower plate is maintained at a higher value than the temperature of the upper one. Rayleigh \cite{Rayleigh1916} found that if the temperature difference is kept at a constant value, then the system possesses a steady-state solution in which there is no motion and convection should take place if this solution becomes unstable. In other words, depending on the temperature difference between the plates, heat can be transferred by conduction or by convection. Assuming variations in only $x_1-x_3$ plane, Saltzman \cite{Saltzman62} considered the equations
\begin{eqnarray}
\begin{array}{l} \label{partial_dif_eqns}
\displaystyle \frac{\partial}{\partial t} \nabla^2 \psi + \frac{\partial \left(\psi, \nabla^2 \psi\right)}{\partial(x_1,x_3)}  - g \varepsilon \frac{\partial \theta}{\partial x_1}-\nu \nabla^4 \psi=0 \\
\displaystyle \frac{\partial \theta}{\partial t} + \frac{\partial \left(\psi, \theta\right)}{\partial(x_1,x_3)} - \frac{\Delta T_0}{H} \frac{\partial \psi}{\partial x_1} -\kappa \nabla^2 \theta=0,
\end{array}
\end{eqnarray}
where $\psi$ is a stream function for the two dimensional motion, $\theta$ is the departure of temperature from that occurring in the state of no convection and the constants $\Delta T_0, H, g, \varepsilon, \nu$ and $\kappa$ denote, respectively, the temperature contrast between the lower and upper boundaries of the fluid, the height of the fluid under consideration, the acceleration of gravity, the coefficient of thermal expansion, the kinematic viscosity and the thermal conductivity \cite{Lorenz63,Saltzman62}. In his study, Saltzman \cite{Saltzman62} achieved an infinite system by means of applying Fourier series methods to system (\ref{partial_dif_eqns}), and then used the simplification procedure proposed by Lorenz \cite{Lorenz60} to obtain a system with finite number of terms.  Lorenz \cite{Lorenz63} set all but three Fourier coefficients equal to zero and as a consequence attained system $(\ref{lorenz_system}),$ which describes an idealized model of a fluid.

In system $(\ref{lorenz_system}),$ the variable $x_1$ is proportional to the circulatory fluid flow velocity, while the variable $x_2$ is proportional to the temperature difference between the ascending and descending currents. Positive $x_1$ values indicate clockwise rotations of the fluid and negative $x_1$ values mean counterclockwise motions. The variable $x_3,$ on the other hand, is proportional to the distortion of the vertical temperature profile from linearity, a positive value indicating that the strongest gradients occur near the boundaries. The parameters $\sigma$ and $r$ are called the Prandtl and Rayleigh numbers, respectively \cite{Alligood,Lorenz63,Sparrow82}.

The dynamics of the Lorenz system (\ref{lorenz_system}) is very rich. For instance, with different values of the parameters $\sigma,$ $r$ and $b,$ the system can exhibit stable periodic orbits, homoclinic explosions, period-doubling bifurcations, and chaotic attractors \cite{Sparrow82}.

The appearance  of chaos in differential/discrete equations may be either endogenous or exogenous. As the first type of chaos birth, one can take into account the irregular motions that occur in Lorenz, R\"{o}ssler, Chua systems, the logistic map, Duffing and Van der Pol equations \cite{Chua86,Dev90,Holmes,Li75,Rossler76,Th02}. To indicate the endogenous irregularity, we use: (i) ingredients of Devaney and Li-Yorke chaos, (ii) period-doubling route to chaos, (iii) intermittency, (iv) positive Lyapunov exponents. Symbolic dynamics and Smale horseshoes have been widely used for that purpose \cite{Dev90,Feigenbaum80,Kennedy01b,Kennedy01,Li75,Pomeau80,Sch05,Wiggins88}. While the endogenous chaos production is widespread and historically unique, the exogenous chaos as generated by irregular perturbations has not been intensively investigated yet. In our study, we will appeal to endogenous chaos, but mostly to exogenous chaos.

In this paper, the main attention is given to the extension of chaos among interconnected Lorenz systems. We make use of unidirectionally coupled Lorenz systems such that the drive system is chaotic and the response system possesses a stable equilibrium or a limit cycle. We theoretically prove that the chaos of the drive system makes the  response system behave also chaotically. Extension of sensitivity and period-doubling cascade are rigorously approved. The appearance of cyclic irregular behavior is discussed, and it is shown that the phenomenon cannot be explained by means of generalized synchronization. Intermittency in coupled Lorenz systems is also demonstrated.

The principal novelty of our investigation is that we create  exogenous chaotic perturbations by means of the solutions of a chaotic Lorenz system, plug it into a regular Lorenz system, and find that chaos is inherited by the solutions of the latter. Such an approach has been widely used for differential equations before, but for regular disturbance functions. That is, it has been shown that an (almost) periodic perturbation function implies the existence of an (almost) periodic solution of the system. While the literature on chaos synchronization \cite{Abarbanel96,Gon04,Hunt97,Kapitaniak94c,Kocarev96,Macau02,Pecora90,Rulkov95} has also produced methods of generating chaos in a system by plugging in terms that are chaotic, it relies on the asymptotic convergence between the chaotic exogenous terms and the solution of the response system for the proof of chaos creation. Instead, we provide a direct verification of the ingredients of chaos for the perturbed system \cite{Akh5}--\cite{Akh12}. Moreover, in Section \ref{butterfly_cyclic} we represent the appearance of cyclic chaos, which cannot be reduced to generalized synchronization. Very interesting examples of applications of discrete dynamics to continuous chaos analysis were provided in the papers \cite{Brown93}--\cite{Brown01}. In these studies, the general technique of dynamical synthesis \cite{Brown93} was developed, and this technique was used in the paper \cite{Akh4}.

There are many published papers which have results about chaos considering first of all its mathematical meaning. This is true either for differential equations \cite{Li12,Wiggins88} or data analysis \cite{Feliks04}. Apparently there are still few articles with meteorological interpretation of chaos ingredients. We suppose that our rigorously approved idea for the extension of chaotic behavior from one Lorenz system to another will give a light for the justification of the erratic behavior observed in dynamical systems of meteorology.

The question \textit{``Does the flap of a butterfly's wings in Brazil set off a tornado in Texas?"} is very impressive and it has done a lot to popularize chaos for both mathematicians and non-mathematicians \cite{Lorenz93}. \textit{From this question one can immediately decide that the butterfly effect is a global phenomenon, and consequently, the underlying mathematics has to be investigated.} Some of the authors say that the question relates sensitive dependence on initial conditions in dynamical systems considered as unpredictability for meteorological observations. Lorenz himself, in successive his talks and the book \cite{Lorenz93}, was obsessed by the question and sincerely believed its possibility. He also supposed that his system can give a key for the positive answer of the question. Generally, analysis of chaotic dynamics in atmospheric models is rather numerical \cite{Fraed86,Itoh96,Kris93,Lorenz05,Mukougawa91} or depend on the observation of time-series \cite{Grassberger83a,Grassberger83b}. In Section \ref{global_weather}, we describe how one can use the rigorously approved results of the present paper to investigate the global behavior of the weather unpredictability. Our suggestions are not about a modelling, but rather an effort to answer the question why the weather is unpredictable at each point of the Earth, on the basis of the Lorenz's meteorological model and other models. We should recognize that all our discussions can be considered as a ``toy object" in the theory, and according to the complexity phenomenon in meteorological investigations one can say that the investigation of chaos in meteorology still remains as a ``toy object" \cite{Grebogi97,Lorenz63}.


\section{Coupled Lorenz Systems}

We couple Lorenz systems unidirectionally in such a way that the existing chaos propagates from one to another. We suppose that the coefficients $\sigma,$ $r$ and $b$ are properly chosen in (\ref{lorenz_system}) so that the system is chaotic. In addition to system (\ref{lorenz_system}), we consider another Lorenz system, 
\begin{eqnarray}
\begin{array}{l} \label{nonperturbed_lorenz_system}
\displaystyle \frac{du_1}{dt}  = - \overline{\sigma} u_1 + \overline{\sigma} u_2 \\
\displaystyle \frac{du_2}{dt}  =  -u_1u_3 +\overline{r} u_1 -u_2  \\
\displaystyle \frac{du_3}{dt}  = u_1u_2-\overline{b}u_3,
\end{array}
\end{eqnarray}
where the parameters $\overline{\sigma},$ $\overline{r}$ and $\overline{b}$ are such that the system is non-chaotic. That is, the system does not possess chaotic motions such that, for example, it admits a global asymptotically stable equilibrium or a globally attracting limit cycle.

In order to realize the chaos extension, we perturb (\ref{nonperturbed_lorenz_system}) with the solutions of (\ref{lorenz_system}) to set up the system
\begin{eqnarray}
\begin{array}{l} \label{perturbed_lorenz_system}
\displaystyle \frac{dy_1}{dt}  = - \overline{\sigma} y_1 + \overline{\sigma} y_2 + g_1(x(t)) \\
\displaystyle \frac{dy_2}{dt}  = -y_1y_3 +\overline{r} y_1 -y_2 + g_2(x(t)) \\
\displaystyle \frac{dy_3}{dt}  = y_1y_2-\overline{b}y_3 + g_3(x(t)),
\end{array}
\end{eqnarray}
where $x(t)=(x_1(t),x_2(t),x_3(t)).$ The conditions on the continuous function $g(x)=(g_1(x),g_2(x),g_3(x))$ is mentioned in the Appendix.

\section{Extension of Sensitivity} \label{Globalization_of_unpredictability}

In this section, we will demonstrate that the divergence of two initially nearby solutions (sensitivity) in the driving chaotic Lorenz system (\ref{lorenz_system}) leads to the presence of the same feature in system (\ref{perturbed_lorenz_system}). Additionally, a third Lorenz system will be considered in order to show the maintainability of the process. The mathematical description of sensitivity and a theoretical proof for its extension are presented in the Appendix.

Let us take into account the system 
\begin{eqnarray}
\begin{array}{l} \label{lorenz_attractor2}
\displaystyle\frac{dy_1}{dt}=-10y_1+10y_2+0.3x_1(t)-0.15 \displaystyle \sin \left(x_1(t)\right) \\
\displaystyle\frac{dy_2}{dt}=-y_1y_3 + 0.35y_1-y_2+ 1.6x_2(t) \\
\displaystyle\frac{dy_3}{dt}=y_1y_2-(8/3)y_3+ 0.1 \displaystyle \tan(x_3(t)/65), 
\end{array}
\end{eqnarray} 
which is in the form of (\ref{perturbed_lorenz_system}) with $\overline{\sigma}=10,$ $\overline r=0.35,$ $\overline b=8/3,$ $g_1(x(t))= 0.3x_1(t)-0.15 \displaystyle \sin \left(x_1(t)\right),$ $g_2(x(t))=1.6x_2(t)$ and $g_3(x(t))=0.1 \displaystyle \tan(x_3(t)/65).$ Here, $x(t)=(x_1(t),x_2(t),x_3(t))$ is a solution of (\ref{lorenz_system}) with $\sigma=10,$ $r=28$ and $b=8/3.$ The coefficients $\sigma,$ $r$ and $b$ are chosen such that chaos takes place in the dynamics of (\ref{lorenz_system}) \cite{Lorenz63}. Besides, system (\ref{nonperturbed_lorenz_system}) with the given values of $\overline{\sigma},$ $\overline r,$ and $\overline b$ possesses a stable equilibrium point \cite{Sparrow82}.  
  
\begin{figure}[ht] 
\centering
\includegraphics[width=6.60cm]{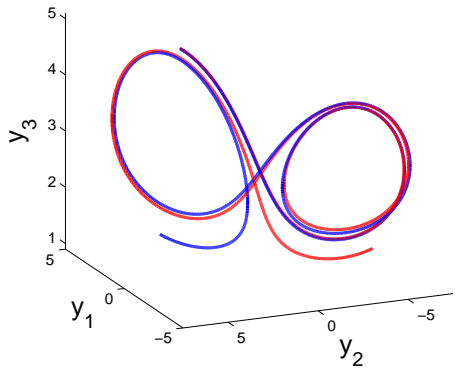}
\caption{Extension of sensitivity in system (\ref{lorenz_attractor2}). The divergence of the initially nearby solutions of system (\ref{lorenz_attractor2}) is observable in the figure.}
\label{lorenz_sensitivity1}
\end{figure}

To reveal numerically the extension of sensitivity in system (\ref{lorenz_attractor2}), we represent in Figure \ref{lorenz_sensitivity1} the projections of two initially nearby trajectories of the unidirectionally coupled system (\ref{lorenz_system})+(\ref{lorenz_attractor2}) on the $y_1-y_2-y_3$ space for $t\in [0,3].$ The trajectory with blue color corresponds to the initial data $x_1(0)=-8.57,$  $x_2(0)=-2.39,$ $x_3(0)=33.08,$ $y_1(0)=3.91,$ $y_2(0)=1.86,$ $y_3(0)=4.39,$ and the one with red color corresponds to the initial data $x_1(0)=-8.53,$ $x_2(0)=-2.47,$ $x_3(0)=33.05,$ $y_1(0)=3.91,$ $y_2(0)=1.87,$ $y_3(0)=4.40.$ The divergence of the initially nearby trajectories seen in Figure \ref{lorenz_sensitivity1} manifests the sensitivity feature in (\ref{lorenz_attractor2}). 

Now, we consider the system
\begin{eqnarray}
\begin{array}{l} \label{lorenz_attractor3}
\displaystyle\frac{dz_1}{dt}=-10z_1+10z_2+12y_1(t) \\
\displaystyle\frac{dz_2}{dt}=-z_1z_3+0.1z_1-z_2+20[y_2(t)+2\displaystyle \arctan(y_2(t)/5)] \\
\displaystyle\frac{dz_3}{dt}=z_1z_2-(8/3)z_3-8y_3(t).
\end{array}
\end{eqnarray}
System (\ref{lorenz_attractor3}) is also in the form of (\ref{perturbed_lorenz_system}), but this time the perturbations $h_1(y(t))=12y_1(t),$ $h_2(y(t))=20[y_2(t)+2\displaystyle \arctan(y_2(t)/5)]$ and $h_3(y(t))=-8y_3(t)$ are provided by the solutions of (\ref{lorenz_attractor2}).

\begin{figure}[ht] 
\centering
\includegraphics[width=6.60cm]{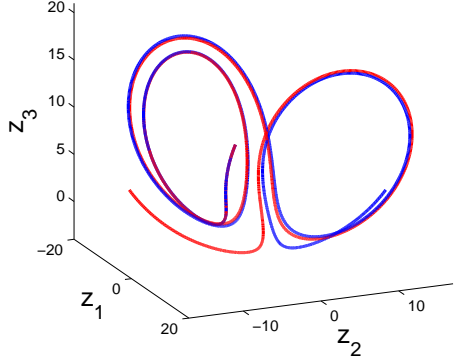}
\caption{Extension of sensitivity in system (\ref{lorenz_attractor3}).}
\label{lorenz_sensitivity2}
\end{figure}

Figure \ref{lorenz_sensitivity2} shows the projections of two trajectories, which are initially nearby, of the $9-$dimensional system (\ref{lorenz_system})+(\ref{lorenz_attractor2})+(\ref{lorenz_attractor3}) on the $z_1-z_2-z_3$ space. The trajectory with blue color has the initial data $x_1(0)=-8.57,$  $x_2(0)=-2.39,$ $x_3(0)=33.08,$ $y_1(0)=3.91,$ $y_2(0)=1.86,$ $y_3(0)=4.39,$ $z_1(0)=6.92,$ $z_2(0)=-6.18,$ $z_3(0)=10.48,$ whereas the one with red color has the initial data $x_1(0)=-8.53,$ $x_2(0)=-2.47,$ $x_3(0)=33.05,$ $y_1(0)=3.91,$ $y_2(0)=1.87,$ $y_3(0)=4.40,$ $z_1(0)=6.89,$ $z_2(0)=-6.18,$ $z_3(0)=10.47.$ The utilized time interval is the same with Figure \ref{lorenz_sensitivity1}. It is seen in Figure \ref{lorenz_sensitivity2} that although the depicted trajectories are initially nearby, later they diverge from each other. In other words, it is demonstrated that the sensitivity of system (\ref{lorenz_attractor2}) is extended to (\ref{lorenz_attractor3}). Moreover, one can conclude from Figure \ref{lorenz_sensitivity1} and Figure \ref{lorenz_sensitivity2} that the system (\ref{lorenz_system})+(\ref{lorenz_attractor2})+(\ref{lorenz_attractor3}) is also sensitive.

\begin{figure}[ht] 
\centering
\includegraphics[width=13.50cm]{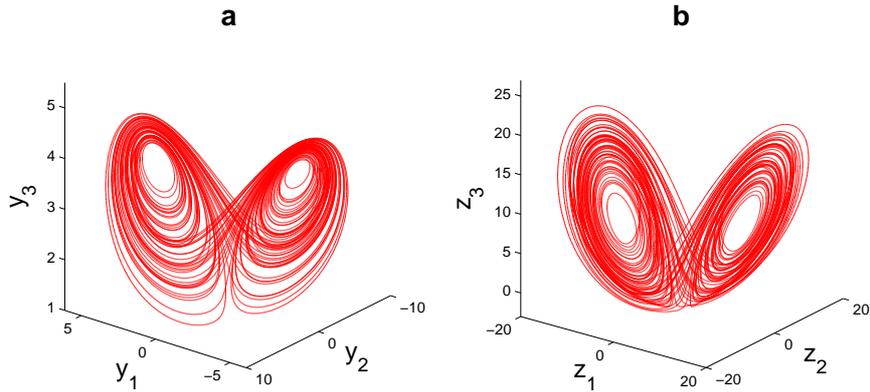}
\caption{The $3-$dimensional projections of the chaotic trajectory of system (\ref{lorenz_system})+(\ref{lorenz_attractor2})+(\ref{lorenz_attractor3}). (a) The projection on the $y_1-y_2-y_3$ space, (b) The projection on the $z_1-z_2-z_3$ space. The picture represented in (a) illustrates the chaotic trajectory of the perturbed Lorenz system (\ref{lorenz_attractor2}), while the picture in (b) corresponds to the perturbed Lorenz system (\ref{lorenz_attractor3}). The pictures represented in (a) and (b) confirm both the extension of chaos and the existence of a chaotic attractor in the $9-$dimensional phase space.}
\label{lorenz_lorenz}
\end{figure}

In the next simulation, the trajectory of system (\ref{lorenz_system})+(\ref{lorenz_attractor2})+(\ref{lorenz_attractor3}) with $x_1(0)=-12.89,$ $x_2(0)=-8.91,$ $x_3(0)=36.59,$ $y_1(0)=-4.21,$ $y_2(0)=-4.96,$ $y_3(0)=3.07,$ $z_1(0)=-14.06,$ $z_2(0)=-8.38,$ $z_3(0)=16.93$ is considered. The three dimensional projections of the trajectory on the $y_1-y_2-y_3$ and $z_1-z_2-z_3$ spaces are depicted in Figure \ref{lorenz_lorenz}, (a) and (b), respectively. Both of the pictures represented in Figure \ref{lorenz_lorenz} manifest not only the chaos extension, but also the existence of a chaotic attractor in the $9-$dimensional phase space. It is worth noting that the projection on the $x_1-x_2-x_3$ space is the classical Lorenz attractor \cite{Lorenz63,Sparrow82}.

The next section is devoted to the extension of chaos obtained through period-doubling cascade.

\section{Extension of Period-Doubling Cascade}

Consider the Lorenz system (\ref{lorenz_system}) in which $\sigma=10,$ $b=8/3$ and $r$ is a parameter \cite{Franceschini80,Sparrow82}. For the values of $r$ between $99.98$ and $100.795$ the system possesses two symmetric stable periodic orbits such that one of them spirals round twice in $x_1>0$ and once in $x_1<0,$ whereas another spirals round twice in $x_1<0$ and once in $x_1>0.$ The book $\cite{Sparrow82}$ calls such periodic orbits as $x^2y$ and $y^2x,$ respectively. More precisely, $``x"$ is written every time when the orbit spirals round in $x_1>0,$ while $``y"$ is written every time when it spirals round in $x_1<0.$ As $r$ decreases towards $99.98$ a period-doubling bifurcation occurs in the system such that two new symmetric stable periodic orbits ($x^2yx^2y$ and $y^2xy^2x$) appear and the previous periodic orbits lose their stability \cite{Franceschini80,Sparrow82}.  According to Franceschini \cite{Franceschini80}, system (\ref{lorenz_system}) undergoes infinitely many period-doubling bifurcations at the parameter values $99.547,$ $99.529,$ $99.5255$ and so on. The sequence of bifurcation parameter values accumulates at $r_{\infty} = 99.524.$ For values of $r$ smaller than $r_{\infty}$ infinitely many unstable periodic orbits take place in the dynamics of (\ref{lorenz_system}) \cite{Franceschini80,Sparrow82}.

To extend the period-doubling cascade of (\ref{lorenz_system}), we take into account the system
\begin{eqnarray}
\begin{array}{l} \label{lorenz_pdc2}
\displaystyle\frac{dy_1}{dt}=-10y_1+10y_2  + 1.8x_1(t) \\
\displaystyle\frac{dy_2}{dt}=-y_1y_3 + 0.27y_1-y_2+ x_2(t) \\
\displaystyle\frac{dy_3}{dt}=y_1y_2-(8/3)y_3+ 0.3x_3(t), 
\end{array}
\end{eqnarray} 
where $x(t)=(x_1(t),x_2(t),x_3(t))$ is a solution of (\ref{lorenz_system}). Note that in the absence of driving, system (\ref{lorenz_pdc2}) admits a stable equilibrium point, i.e., system (\ref{nonperturbed_lorenz_system}) with $\overline{\sigma}=10,$ $\overline{r}=0.27$ and $\overline{b}=8/3$ does not admit chaos. 

By using Theorem $15.8$ \cite{Yoshizawa75}, one can verify that for each periodic $x(t),$ there exists a periodic solution of (\ref{lorenz_pdc2}) with the same period.
 
\begin{figure}[ht] 
\centering
\includegraphics[width=13.5cm]{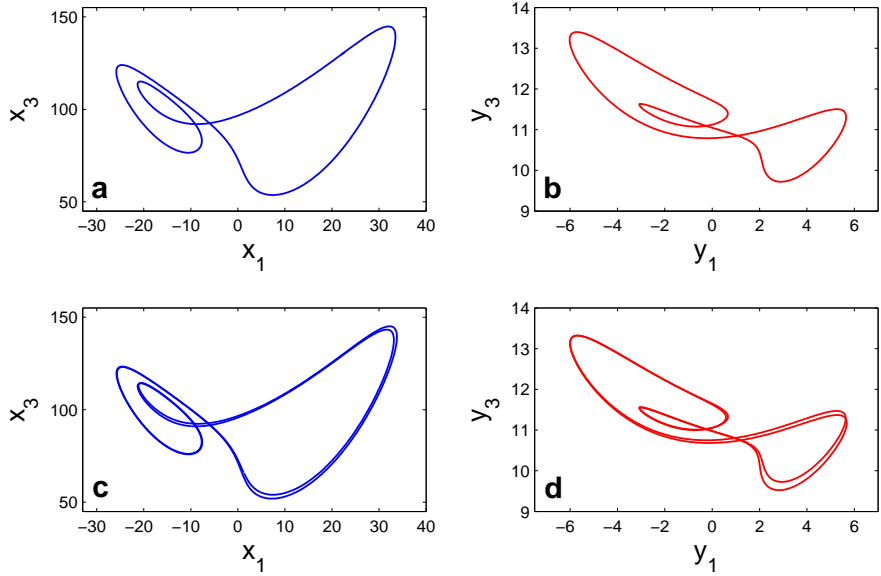}
\caption{The stable periodic orbits of the systems (\ref{lorenz_system}) and (\ref{lorenz_pdc2}).}
\label{pdc1}
\end{figure}

In Figure \ref{pdc1}, we illustrate the stable periodic orbits of systems (\ref{lorenz_system}) and (\ref{lorenz_pdc2}). Figure \ref{pdc1}, (a) shows the $y^2x$ periodic orbit of (\ref{lorenz_system}) for $r=100.36,$ while Figure \ref{pdc1}, (b) depicts the corresponding periodic orbit of system (\ref{lorenz_pdc2}). Similarly, Figure \ref{pdc1}, (c) and (d) represent the $y^2xy^2x$ periodic orbit of (\ref{lorenz_system}) with $r=99.74$ and the corresponding periodic orbit of (\ref{lorenz_pdc2}), respectively. Figure \ref{pdc1} confirms that if (\ref{lorenz_system}) has a periodic orbit, then (\ref{lorenz_pdc2}) also has a periodic orbit with the same period.  

\begin{figure}[ht] 
\centering
\includegraphics[width=13.5cm]{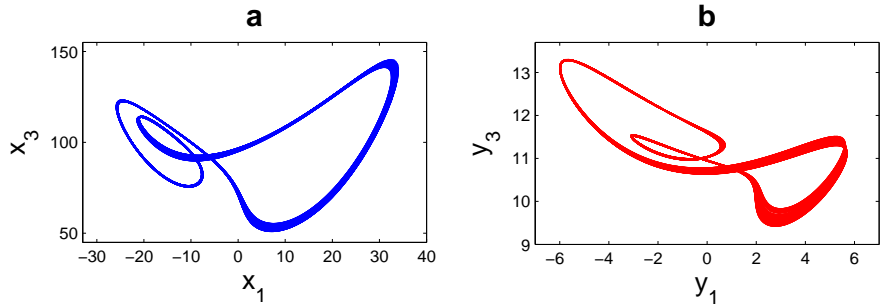}
\caption{Extension of period-doubling cascade in the unidirectionally coupled Lorenz systems $(\ref{lorenz_system})+(\ref{lorenz_pdc2}).$}
\label{pdc2}
\end{figure}

\begin{figure}[ht] 
\centering
\includegraphics[width=12.5cm]{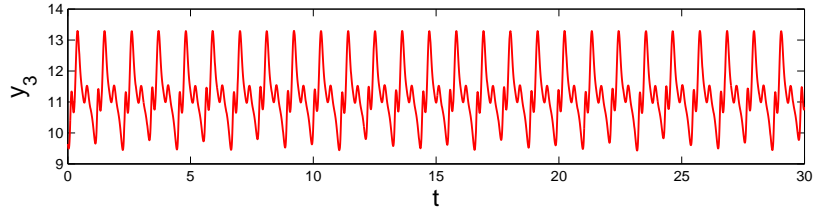}
\caption{The irregular behavior of the $y_3$ coordinate of system (\ref{lorenz_pdc2}) with $r=99.51.$}
\label{pdc3}
\end{figure}

Next, we continue with the extension of period-doubling cascade in Figure \ref{pdc2}. The projection of the trajectory of system (\ref{lorenz_system}) with $r=99.51$ corresponding to the initial data $x_1(0)=10.58,$ $x_2(0)=28.19,$ $x_3(0)=53.32$ on the $x_1-x_3$ plane is shown in Figure \ref{pdc2}, (a). Making use of the initial data $y_1(0)=2.23,$ $y_2(0)=1.26,$  $y_3(0)=9.64,$ the projection of the corresponding trajectory of (\ref{lorenz_pdc2}) on the $y_1-y_3$ plane is depicted in Figure \ref{pdc2}, (b). Moreover, the irregular behavior of the $y_3$ coordinate over time is illustrated in Figure \ref{pdc3}. The simulation results reveal that the period-doubling cascade of (\ref{lorenz_system}) is extended to (\ref{lorenz_pdc2}). A theoretical investigation of the extension of period-doubling cascade is provided in the Appendix.
 
In the next section, the extension of intermittency is considered.


\section{Extension of Intermittency}

Pomeau and Manneville \cite{Pomeau80} observed intermittency in the Lorenz system (\ref{lorenz_system}), where $\sigma =10,$ $b=8/3$ and $r$ is slightly larger than the critical value $r_c\approx166.06.$  
Let us use $r=166.3$ in system (\ref{lorenz_system}) such that intermittency is present. We perturb system (\ref{nonperturbed_lorenz_system}), where $\overline{\sigma}=10,$ $\overline{r}=0.35,$ $\overline{b}=8/3,$ with solutions of (\ref{lorenz_system}), and set up the following system, 
\begin{eqnarray}
\begin{array}{l} \label{lorenz_intermittency}
\displaystyle\frac{dy_1}{dt}=-10 y_1 + 10 y_2+0.7x_1(t) \\
\displaystyle\frac{dy_2}{dt}= -y_1y_3 +0.35 y_1 -y_2-x_2(t) \\
\displaystyle\frac{dy_3}{dt}=y_1y_2-(8/3)y_3+0.2x_3(t).
\end{array}
\end{eqnarray}

\begin{figure}[ht] 
\centering
\includegraphics[width=12.00cm]{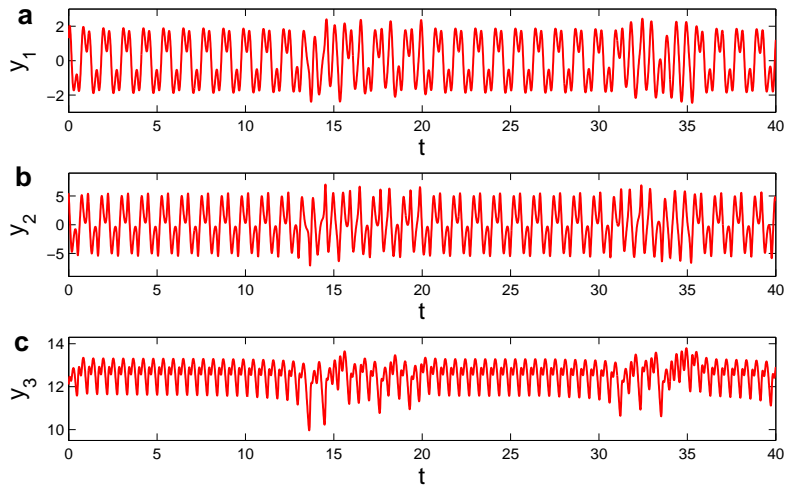}
\caption{Extension of intermittency in system (\ref{lorenz_intermittency}). The behaviors of the $y_1, y_2$ and $y_3$ coordinates are shown in pictures (a), (b) and (c), respectively. The extension of the intermittent behavior is observable such that regular motions are interrupted by irregular ones.}
\label{inter2} 
\end{figure}

The graphs of the $y_1,y_2$ and $y_3$ coordinates of (\ref{lorenz_intermittency}) are shown in Figure \ref{inter2} by making use of the initial data $x_1(0)=-23.3,$ $x_2(0)=38.3,$ $x_3(0)=193.4,$ $y_1(0)=1.3,$ $y_2(0)=5.5,$ $y_3(0)=12.1.$  It is revealed in Figure \ref{inter2} that regular oscillations are interrupted by irregular ones, i.e., the intermittent behavior of the prior Lorenz system is extended even if system (\ref{nonperturbed_lorenz_system}) admits a stable equilibrium point.

For some parameter values, the Lorenz system can exhibit limit cycles \cite{Sparrow82}. In the next section, we will consider the Lorenz system (\ref{nonperturbed_lorenz_system}) with a globally attracting limit cycle and verify numerically how to achieve a motion that behaves chaotically and cyclically in the same time.

\section{Cyclic Chaos} \label{butterfly_cyclic}

In our previous illustrations, we considered system (\ref{nonperturbed_lorenz_system}) with a stable equilibrium point. Now, we consider the model with a limit cycle. The numerical simulations represented in this section are theoretically based on the paper \cite{Akh12}, where the main result is about the existence of infinitely many unstable periodic solutions and extension of sensitivity.

Let us consider the systems (\ref{lorenz_system}) and (\ref{nonperturbed_lorenz_system}) with the coefficients  $\sigma=10,$ $r-28,$ $b=8/3$ and $\overline{\sigma}=10,$ $\overline{r}=350,$ $\overline{b}=8/3,$ respectively, such that (\ref{lorenz_system}) is chaotic and (\ref{nonperturbed_lorenz_system})  possesses a globally attracting limit cycle \cite{Sparrow82}.
We perturb system (\ref{nonperturbed_lorenz_system}) with the solutions of (\ref{lorenz_system}), and constitute the system
\begin{eqnarray}
\begin{array}{l} \label{lorenz_around_cycle}
\displaystyle\frac{dy_1}{dt}=-10 y_1 + 10 y_2+2.3x_1(t) \\
\displaystyle\frac{dy_2}{dt}= -y_1y_3 +350 y_1 -y_2+2x_2(t)\\
\displaystyle\frac{dy_3}{dt}=y_1y_2-(8/3)y_3+1.5x_3(t).
\end{array}
\end{eqnarray}

\begin{figure}[ht] 
\centering
\includegraphics[width=13.00cm]{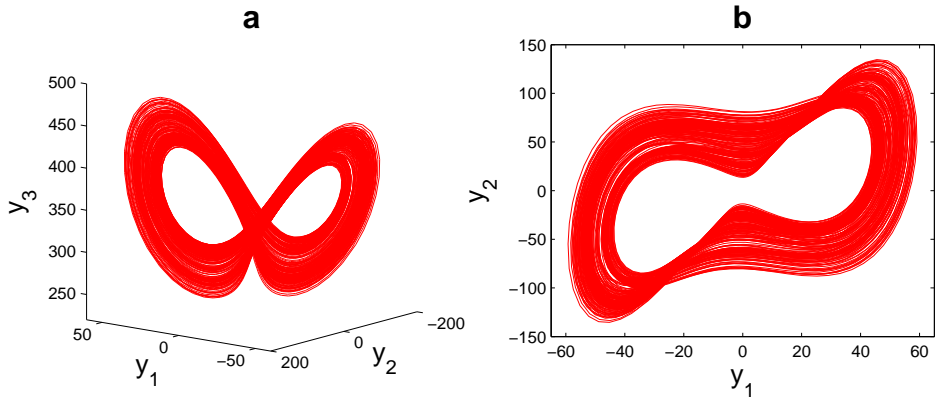}
\caption{The projections of the chaotic trajectory produced by the coupled system $(\ref{lorenz_system})+(\ref{lorenz_around_cycle}).$ (a) The $3-$dimensional projection on the $y_1-y_2-y_3$ space; (b) The $2-$dimensional projection on the $y_1-y_2$ plane. The pictures in (a) and (b) represent a motion that behaves both chaotically and cyclically around the stable limit cycle of (\ref{nonperturbed_lorenz_system}).}
\label{lorenz_cycle}
\end{figure}

Making use of the solution of (\ref{lorenz_system}) corresponding to the initial data $x_1(0)=5.71,$  $x_2(0)=9.01$, $x_3(0)=17.06,$ we depict the trajectory of (\ref{lorenz_around_cycle}) with $y_1(0)=-21.67,$ $y_2(0)=34.33,$ $y_3(0)=346.38$ in Figure \ref{lorenz_cycle}, (a). The projection of the same trajectory on the $y_1-y_2$ plane is shown in Figure \ref{lorenz_cycle}, (b). It is seen in both figures that the trajectory behaves chaotically around the limit cycle of (\ref{nonperturbed_lorenz_system}).

\begin{figure}[ht] 
\centering
\includegraphics[width=13.00cm]{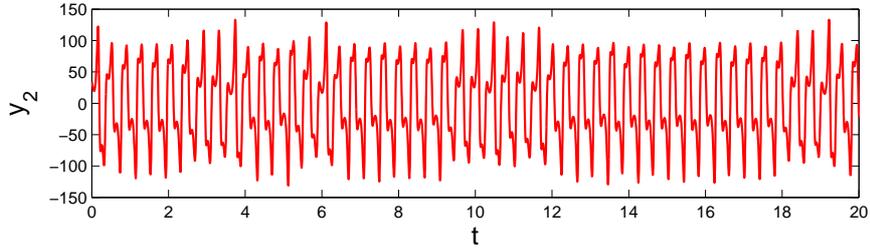}
\caption{The time-series for the $y_2-$coordinate of system $(\ref{lorenz_around_cycle}).$ The picture confirms the chaotic behavior of the motion. The remaining coordinates of system $(\ref{lorenz_around_cycle}),$ which are not just pictured here, behave also chaotically.}
\label{cycle_timeseries}
\end{figure}

To confirm one more time that the trajectory considered in Figure \ref{lorenz_cycle} is essentially chaotic, the graph of the $y_2$ coordinate of system (\ref{lorenz_around_cycle}) is illustrated in Figure \ref{cycle_timeseries}. Although system (\ref{nonperturbed_lorenz_system}) possesses a globally attracting limit cycle, the simulations seen in Figure \ref{lorenz_cycle} and Figure \ref{cycle_timeseries} indicate that the applied perturbation makes the system (\ref{lorenz_around_cycle}) behave chaotically. In other words, the chaotic behavior is seized by the limit cycle of system (\ref{nonperturbed_lorenz_system}), and as a result a motion which behaves both chaotically and cyclically appears.

In order to compare our approach with that of generalized synchronization \cite{Abarbanel96,Gon04,Hunt97,Kocarev96,Rulkov95}, let us apply the auxiliary system approach \cite{Abarbanel96,Gon04} to the couple $(\ref{lorenz_system})+(\ref{lorenz_around_cycle}).$

The corresponding auxiliary system is
\begin{eqnarray}
\begin{array}{l} \label{lorenz_around_cycle_auxiliary}
\displaystyle\frac{dz_1}{dt}=-10 z_1 + 10 z_2+2.3x_1(t) \\
\displaystyle\frac{dz_2}{dt}= -z_1z_3 +350 z_1 -y_2+2x_2(t)\\
\displaystyle\frac{dz_3}{dt}=z_1z_2-(8/3)z_3+1.5x_3(t).
\end{array}
\end{eqnarray}

\begin{figure}[ht] 
\centering
\includegraphics[width=8.00cm]{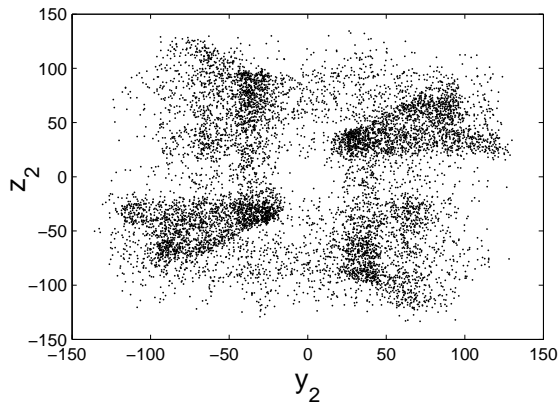}
\caption{Application of the auxiliary system approach to system $(\ref{lorenz_system})+(\ref{lorenz_around_cycle})$ indicates that generalized synchronization does not exist for the couple.}
\label{lorenz_cycle_gs}
\end{figure}

The projection of the stroboscopic plot of the $9-$dimensional system $(\ref{lorenz_system})+(\ref{lorenz_around_cycle})+(\ref{lorenz_around_cycle_auxiliary})$ on the $y_2-z_2$ plane is depicted in Figure \ref{lorenz_cycle_gs}. The figure is obtained by marking the trajectory with the initial data $x_1(0) =5.71,$ $x_2(0) =9.01,$ $x_3(0)=17.06,$ $y_1(0) =-21.67 ,$ $y_2(0) =34.33,$ $y_3(0)=346.38,$ $z_1(0) =-46.26,$ $z_2(0) = -49.73 ,$ $z_3(0)=415.87,$  and by omitting the first $200$ iterations. It is observable in Figure \ref{lorenz_cycle_gs} that the stroboscopic plot is not on the line $z_2 = y_2.$ Therefore, we conclude that generalized synchronization does not take place in the dynamics of the couple $(\ref{lorenz_system})+(\ref{lorenz_around_cycle}).$
 
Another approach to investigate the presence or absence of generalized synchronization is the evaluation of conditional Lyapunov exponents \cite{Gon04,Kocarev96,Pecora90}. 

To determine the conditional Lyapunov exponents, we take into account the following variational equations for system (\ref{lorenz_around_cycle}),
\begin{eqnarray}
\begin{array}{l} \label{lorenz_cond_lyp_exp}
\displaystyle\frac{d \xi_1}{dt}=-10 \xi_1 + 10 \xi_2  \\
\displaystyle\frac{d \xi_2}{dt}= (-y_3(t) +350) \xi_1 -\xi_2 -y_1(t) \xi_3\\
\displaystyle\frac{d \xi_3}{dt}= y_2(t)\xi_1 +y_1(t)\xi_2 -(8/3)\xi_3.
\end{array}
\end{eqnarray}
Utilizing the solution $y(t)$ of (\ref{lorenz_around_cycle}) corresponding to the initial data $x_1(0) =5.71,$ $x_2(0) =9.01,$ $x_3(0)=17.06,$ $y_1(0) =-21.67,$ $y_2(0) =34.33,$ $y_3(0)=346.38,$ the largest Lyapunov exponent of system (\ref{lorenz_cond_lyp_exp}) is evaluated as $0.0226.$ That is, system (\ref{lorenz_around_cycle}) possesses a positive conditional Lyapunov exponent, and this result reveals one more time the absence of generalized synchronization.


\section{Self-organization and Synergetics in Lorenz Systems}\label{synergetics}

To illustrate the extension of chaos in large collections of interconnected Lorenz systems, let us introduce the following $27-$dimensional system consisting of the subsystems $S_1,S_2,\ldots,S_9:$ 

\begin{equation*}  
\left.
\begin{aligned}
\displaystyle \frac{dx_1}{dt} &= -10x_1+10x_2 \\
\displaystyle \frac{dx_2}{dt} &= -x_1 x_3+28x_1-x_2 \\ 
\displaystyle \frac{dx_3}{dt} &= x_1 x_2-(8/3) x_3 \\
\end{aligned}
\right\} S_1 
\end{equation*}
\begin{equation*}
\left.
\begin{aligned}
\displaystyle \frac{dy_1}{dt} &= -10 y_1+10 y_2 +8 x_1(t)  \\
\displaystyle \frac{dy_2}{dt} &= -y_1 y_3+0.21 y_1-y_2 + x_2(t) +0.001 x_2^3(t) \\
\displaystyle \frac{dy_3}{dt} &= y_1 y_2-(8/3)y_3 +2 x_3(t) \\
\end{aligned}
\right\}S_2 
\end{equation*}
\begin{equation*}
\left.
\begin{aligned}
\displaystyle \frac{dz_1}{dt} &= -10 z_1 +10 z_2 +4 x_2(t) \\
\displaystyle \frac{dz_2}{dt} &= -z_1 z_3+0.02 z_1-z_2 +3 x_3(t) \\
\displaystyle \frac{dz_3}{dt} &= z_1 z_2-(8/3) z_3 + x_1(t) + 0.1 \cos(x_1(t)) \\
\end{aligned}
\right\}S_3 
\end{equation*}
\begin{equation*}
\left.
\begin{aligned}
\displaystyle \frac{dw_1}{dt} &= -10w_1+10w_2 -x_1(t)\\
\displaystyle \frac{dw_2}{dt} &= -w_1 w_3+0.34w_1-w_2 +4\tanh(x_2(t)) \\ 
\displaystyle \frac{dw_3}{dt} &= w_1 w_2-(8/3)w_3 -5 x_3(t) \\
\end{aligned}
\right\}S_4 
\end{equation*}
\begin{equation*}
\left.
\begin{aligned}
\displaystyle \frac{d\zeta_1}{dt} &= -10 \zeta_1 +10 \zeta_2 +\tan(y_1(t)/20)  \\ 
\displaystyle \frac{d\zeta_2}{dt} &=-\zeta_1 \zeta_3+ 0.12 \zeta_1-\zeta_2 +2.5y_2(t) \\
\displaystyle \frac{d\zeta_3}{dt} &= \zeta_1  \zeta_2-(8/3) \zeta_3 -10y_3(t) \\
\end{aligned}
\right\}S_5 
\end{equation*}
\begin{equation*}
\left.
\begin{aligned}
\displaystyle \frac{d\eta_1}{dt} &=-10\eta_1+10\eta_2 +8y_1(t)  \\
\displaystyle \frac{d\eta_2}{dt} &=-\eta_1 \eta_3+0.29\eta_1-\eta_2 +4.5y_3(t) \\
\displaystyle \frac{d\eta_3}{dt} &=\eta_1 \eta_2-(8/3)\eta_3 -e^{y_2(t)/30} \\
\end{aligned}
\right\}S_6
\end{equation*}
\begin{equation*} 
\left.
\begin{aligned}
\displaystyle \frac{d\kappa_1}{dt} &=-10 \kappa_1+10 \kappa_2 +4z_1(t) \\
\displaystyle \frac{d\kappa_2}{dt} &=-\kappa_1\kappa_3+0.19 \kappa_1-\kappa_2 +9z_2(t) \\
\displaystyle \frac{d\kappa_3}{dt} &=\kappa_1 \kappa_2-(8/3)\kappa_3 +6z_3(t) \\
\end{aligned}
\right\}S_7 
\end{equation*}
\begin{equation*}
\left.
\begin{aligned}
\displaystyle \frac{d\rho_1}{dt} &= -10 \rho_1 +10 \rho_2 +4 w_1(t)  \\
\displaystyle \frac{d\rho_2}{dt} &= -\rho_1 \rho_3+0.17 \rho_1-\rho_2+7 w_2(t) \\
\displaystyle \frac{d\rho_3}{dt} &= \rho_1\rho_2-(8/3)\rho_3 -3\tanh(w_3(t)) \\
\end{aligned}
\right\}S_8 
\end{equation*}
\begin{equation*}
\left.
\begin{aligned}
\displaystyle \frac{d\tau_1}{dt} &=-10\tau_1+10\tau_2+\arctan(w_1(t))  \\
\displaystyle \frac{d\tau_2}{dt} &=-\tau_1\tau_3+0.32\tau_1-\tau_2 + 9 w_2(t) \\ 
\displaystyle \frac{d\tau_3}{dt} &=\tau_1\tau_2-(8/3)\tau_3 + w_3(t). 
\end{aligned}
\right\}S_9
\end{equation*}
 
\begin{figure}[ht] 
\centering
\includegraphics[width=4.4cm]{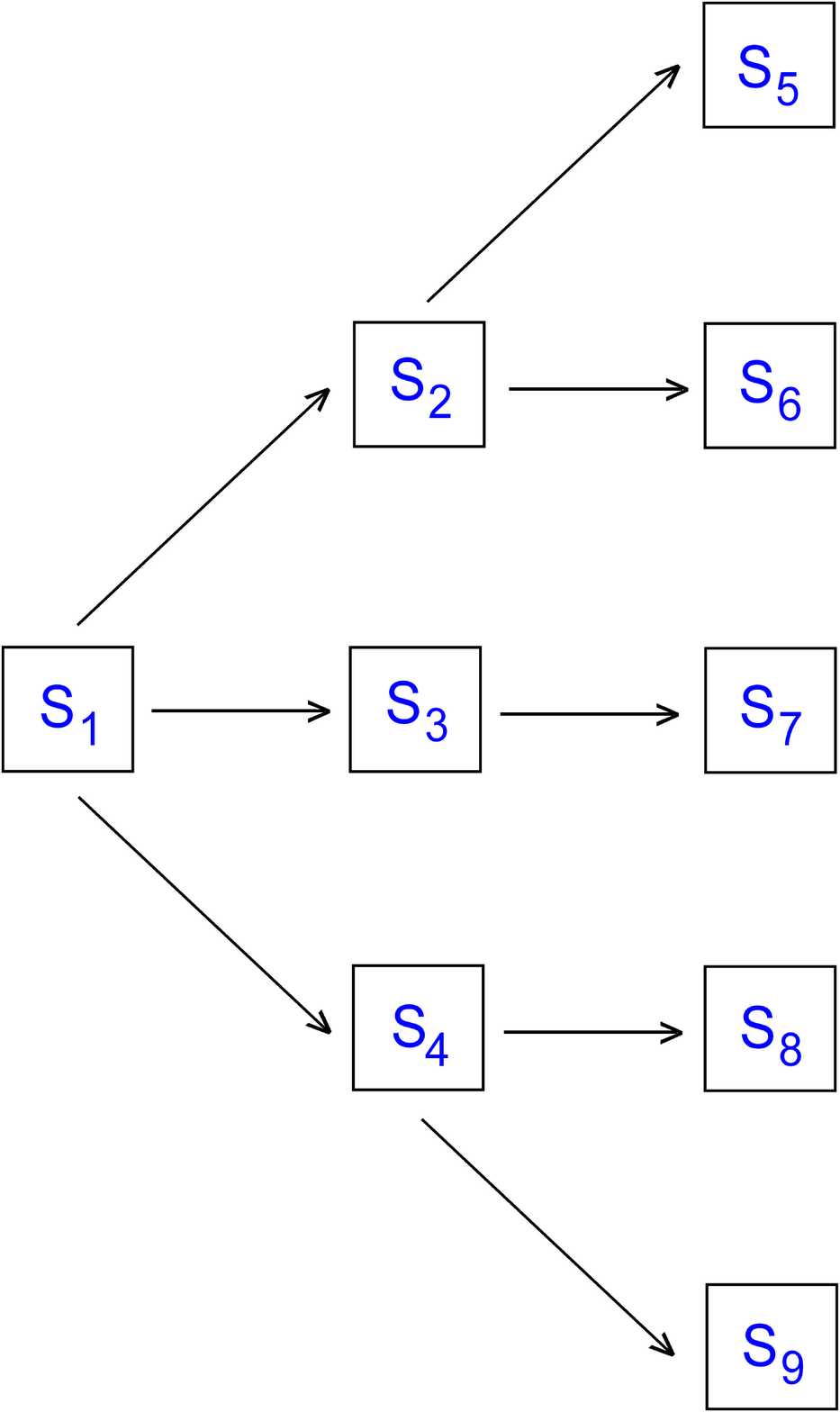}
\caption{The topology of the chaos extension through the interconnected Lorenz systems $S_1,S_2,\ldots,S_9.$ }
\label{intro_systems1}
\end{figure}

\begin{figure}[ht] 
\centering
\includegraphics[width=8.0cm]{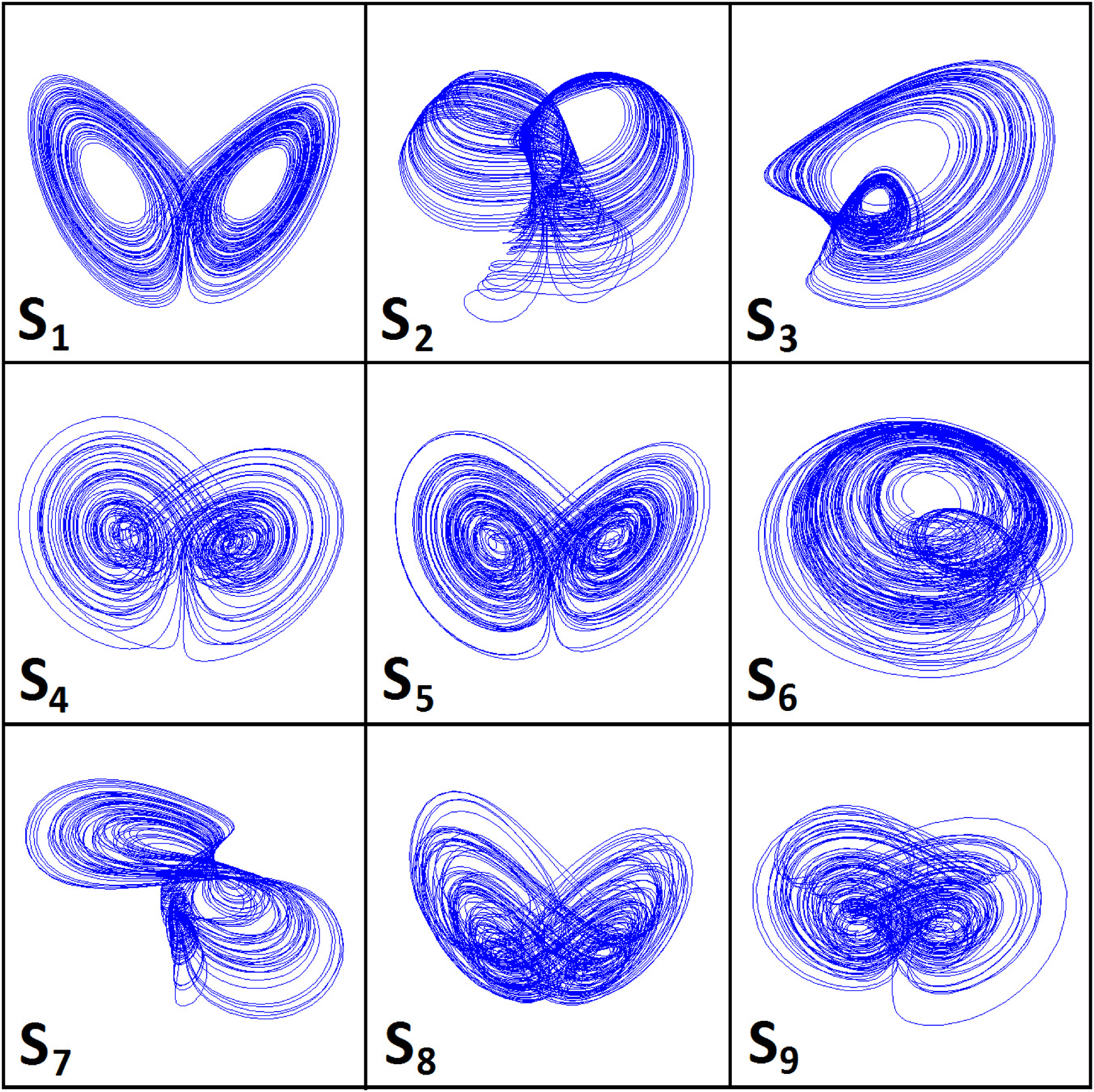}
\caption{The chaotic attractors of the Lorenz systems $S_1,S_2,\ldots,S_9.$}
\label{intro_systems2}
\end{figure}

The coefficients of $S_1$ are chosen in such a way that the system is chaotic \cite{Lorenz63}. The systems $S_2,S_3,\ldots,S_9$ are designed such that if the corresponding perturbations $x(t),$ $y(t),$ $z(t),$ $w(t)$ are chaotic, then the systems possess chaos. However, in the absence of the perturbations, $S_2,S_3,\ldots,S_9$ admit stable equilibria and they are all non-chaotic. The connection topology of the systems $S_1,S_2,\ldots,S_9$ is represented in Figure \ref{intro_systems1}. On the other hand, Figure \ref{intro_systems2} depicts the chaotic attractors corresponding the each $S_i,$ $i=1,2,\ldots,9,$ such that collectively the picture can be considered as the chaotic attractor of the whole $27-$dimensional system. One can confirm that Figure \ref{intro_systems2} supports our ideas such that the chaos of $S_1$ generates chaos in the remaining subsystems even if they are non-chaotic in the absence of the perturbations.

The idea of the transition of chaos from one system to another as well as the arrangement of chaos in an ordered way can be considered as another level of self-organization \cite{haken83,nicolis}. Durrenmatt \cite{durrenmatt} described that ``... a system is self-organizing if it acquires a spatial, temporal or functional structure without specific interference from the outside. By ``specific" we mean that the structure of functioning is not impressed on the system, but the system is acted upon from the outside in a nonspecific fashion." There are three approaches to self-organization, namely thermodynamic (dissipative structures), synergetic and the autowave. For the theory of dynamical systems (e.g. differential equations) the phenomenon means that an autonomous system of equations admits a regular and stable motion (periodic, quasiperiodic, almost periodic). In the literature, this is called as autowaves processes \cite{avk} or self-excited oscillations \cite{Moon}. We are inclined to add to the list another phenomenon, which is a consequence of the chaos extension. Consider the subsystems $S_1, S_2, \ldots, S_9$ once again (in general, arbitrary finite number of systems can be considered). Because of the connections and the conditions discovered in our analysis (see Appendix), likewise $S_1$ all the other subsystems, $S_i,$ $i=2,3,\ldots,9,$ are also chaotic. We suppose that this is a self-organization. This phenomenon can be restricted only for autonomous systems or it can be even interpreted for non-autonomous systems, too. So, we can say that the extension of unpredictability is an example of self-organization, that is a coherent behavior of a large number of systems \cite{haken83}.

In his fascinating paper, the German theoretical physicist Hermann Haken \cite{haken83} introduced a new interdisciplinary  field of science, synergetics, which deals with the origins and the evolution of spatio-temporal structures. The profound part of synergetics is based on the dynamical systems theory. Depending on the discussion of our manuscript, it is natural that we concentrate on the differential equations, and everything that will be mentioned below about synergetics concerns first of all dynamical systems with mathematical approach. One of the main features of systems in synergetics is self-organization, which has been discussed above, and we approved that the phenomenon is present in the extension of chaos among Lorenz systems. According to Haken \cite{haken83}, the central question in synergetics is whether there are general principles that govern the self-organized formation of structures and/or functions. The main principles by the founder of the theory are instability, order parameters, and slaving \cite{haken83}. Instability is understood as the formation or collapse of structures (patterns). This is very common in fluid dynamics, lasers, chemistry and biology \cite{haken83,haken88,haken02,murray,vorontsov}. A number of examples of instability can be found in the literature about morphogenesis  \cite{Turing} and the pattern formation examples can be found in fluid dynamics. The phenomenon is called as instability because the former state of fluid transforms to a new one, loses its ability to persist, and becomes unstable. We see instability in the chaos extension, as consecutive chaotification of systems $S_2,S_3,\ldots,S_9$ joined to the source $S_1$ of chaos. The concepts of the order parameter and slaving are strongly connected in synergetics. For differential equations theory, order parameters mean those phase variables, whose behavior formate the main properties of a macroscopic structure, which dominate over all other variables in the formation such that they can even depend on the order parameters functionally. The dependence that is proved (discovered) mathematically is what we call as slaving. It is not difficult to see in the chaos extension mechanism that the variables of the system $S_1$ are order parameters, and they determine the chaotic behavior of the joined systems' variables. That is, the slaving principle is present here.

The next section is devoted to the possible connections of our results about interconnected Lorenz systems with the global weather unpredictability. 

\section{Connection with Global Weather Unpredictability} \label{global_weather}

Lorenz \cite{Lorenz63} was the first who discovered sensitivity with the aid of system (\ref{lorenz_system}) and then made the conclusion on the butterfly effect. Nowadays, there is an agreement that the butterfly effect exists, if we mean \textit{sensitivity=unpredictability}. It seems that Lorenz himself believed that sensitivity discovered in his equation is a strong indicator of the butterfly effect in its  original meteorological sense. Possibly his intuition is based on the idea that the system of ordinary differential equations is derived from a system of partial differential equations. There should be a deeper interpretation for the effect of chaotic dynamics in the three dimensional system on the infinite dimensional one. We also believe that the opinion of Lorenz, who considered his results as an evidence of the meteorological butterfly effect, is very reasonable. Moreover, his claim has to be considered as a challenging problem for mathematicians. If one thinks positively on the subject, then by our opinion several next questions emerge. The first one is whether sensitivity in meteorological models is a reflection of the butterfly effect. Definitely, this question needs a thoroughly investigation. Possibly it requests a deep analysis on the basis of ordinary and partial differential equations. The problem is not solved in this section at all. We axiomatize somehow the state assuming that the butterfly effect is sensitivity in the mathematical sense or, more generally, chaos. We can also reduce the question by considering the problem of unpredictability through sensitivity. Consequently, the following questions are reasonable: Can one explain the \textit{global} unpredictability of weather by applying models similar to the Lorenz system? How can Lorenz systems be utilized for a {\it global} description of the weather? 

The physical properties of the atmosphere are not the same throughout the Earth. The tropical atmosphere possesses considerably different behavior from those in the temperate and polar latitudes, as if it were a different fluid \cite{Lorenz93}. Taking inspiration from its multifaceted structure, we propose to consider the atmosphere divided into subregions such that the dynamic properties of each region differs considerably from the others. In this case, one can suppose that the dynamics of each subregion of the atmosphere subjects to its own Lorenz system. That is, for different subregions the coefficients of the corresponding Lorenz system are different. Since for some parameter values chaos can take place in the Lorenz system and for some not, such chaotic or non-chaotic motions should have prolonged forever, conflicting the realistic dynamics of the atmosphere, where global unpredictability is present. To extend our attitude for the butterfly effect, we propose that instability, which may occur in a subregion, can be imported to neighbor subregions of the atmosphere, such that chaos occurs not only endogenously, but also exogenously. In other words, exterior perturbations influencing a part of the atmosphere may cause a chaotic behavior to occur in that region. In addition to this, we suppose that these perturbations most probably originate through the neighboring regions within the atmosphere, and the dynamics of coupled Lorenz systems can help to analyze this.

The results presented in the previous sections can be useful to investigate the underlying reasons of the global weather unpredictability under the following assumptions:
\begin{enumerate}
\item[(i)] The whole atmosphere of the Earth is partitioned in a finite number of subregions;
\item[(ii)] In each of the subregions the dynamics of the weather is governed by the Lorenz system with certain coefficients;
\item[(iii)] There are subregions for which the corresponding Lorenz systems admit a chaos with the main ingredient as sensitivity, which means unpredictability of weather in the meteorological sense, and there are subregions, where Lorenz systems are non-chaotic and with equilibriums or cycles as global attractors;    
\item[(iv)] The Lorenz systems are connected unidirectionally.
\end{enumerate}

Let us localize the global process by taking into account only two adjacent subregions of the atmosphere, labeled A and B. In the beginning, the subregion A is assumed to be chaotic, while the subregion B is non-chaotic. By the phrase ``chaotic subregion" we mean that the coefficients of the corresponding Lorenz system are such that the system possesses a chaotic attractor. In a similar way, one should understand from the phrase ``non-chaotic subregion" that the corresponding Lorenz system does not exhibit chaotic motions such that, for example, it admits a global asymptotically stable equilibrium or a globally attracting limit cycle. 

\begin{figure}[ht] 
\centering
\includegraphics[width=6.5cm]{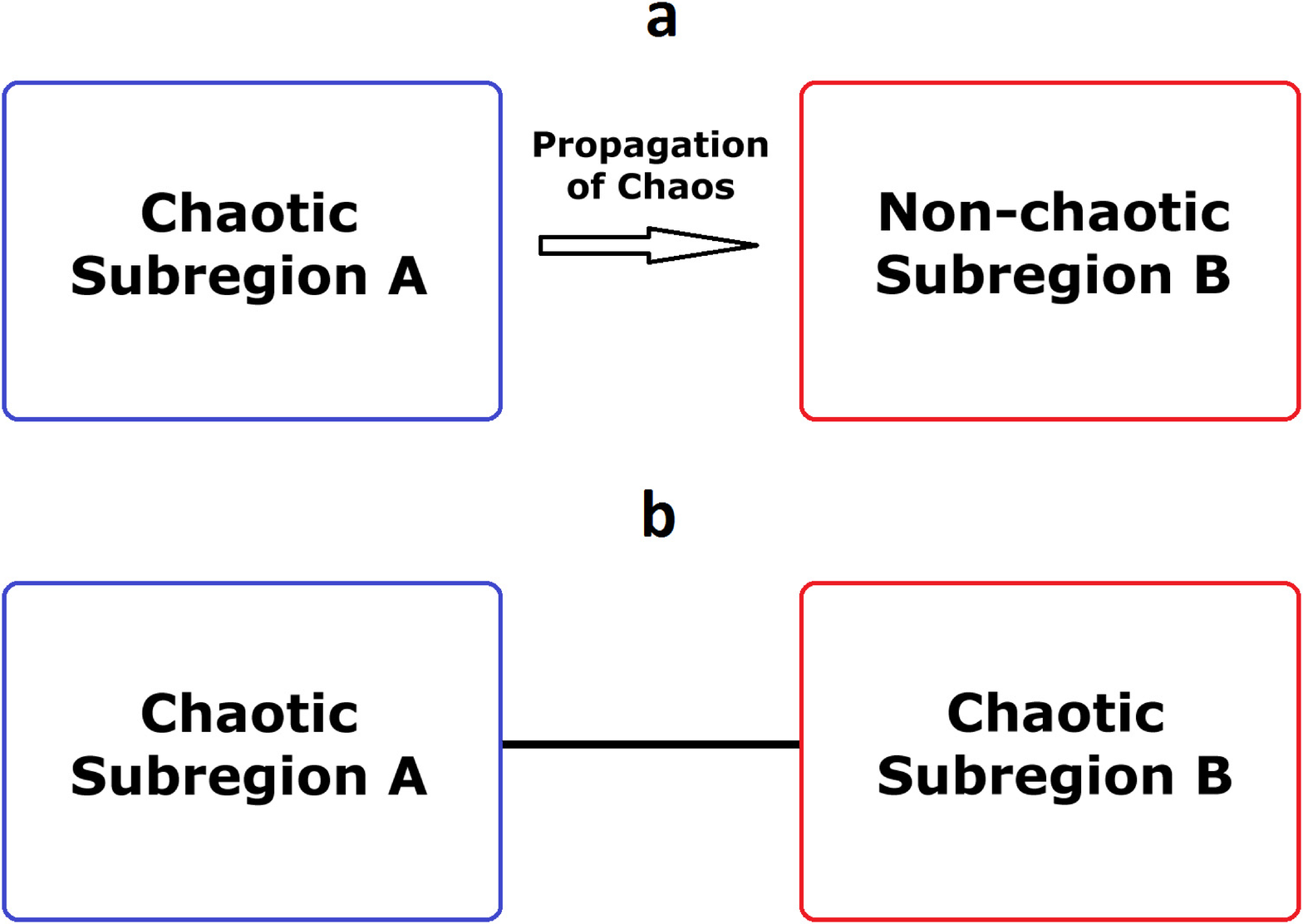}
\caption{Schematic representation of the chaos extension mechanism. (a) The dynamics during the transmission of chaos, (b) The state of the weather after the transmission of chaos. The unidirectional coupling of Lorenz systems gives rise to the chaotification of the initially non-chaotic system such that as a result the unpredictability has been propagated from subregion A to subregion B.}
\label{lorenz_fig}
\end{figure}

Suppose that the dynamics of A is described by the chaotic Lorenz system (\ref{lorenz_system}). Besides, system (\ref{perturbed_lorenz_system}) represents the dynamics of B after the transmission of chaos, and system (\ref{nonperturbed_lorenz_system}) represents the dynamics of B before the process is carried out. According to the theoretical results of the present paper, the chaos of the subregion A influences the subregion B in such a way that the latter also becomes chaotic even if it is initially non-chaotic. The propagation mechanism is represented schematically in Figure \ref{lorenz_fig}. By chaos propagation, we mean the process of unidirectional coupling of Lorenz systems. Figure \ref{lorenz_fig}, (a) illustrates the dynamics during the transmission of chaos. After the transmission of unpredictability is achieved, the dynamics of both subregions, A and B, exhibit chaotic behavior as shown in Figure \ref{lorenz_fig}, (b).

The mentioned local process can be maintained by considering more subregions, whose dynamics are also described by Lorenz systems. For instance, one can suppose that the systems $S_1, S_2,\ldots,S_9$ defined in Section \ref{synergetics} reflect the dynamics of the weather in nine different subregions. The applied perturbations may not be in accordance with realistic air flows in the atmosphere. However, the exemplification reveals the propagation of unpredictability and indicate the possibility for the usage of different types of perturbations in the systems. We do not take into account changes which may happen because of the day light evolution, variety of land forms, seasonal differences in the region etc., but what we propose is to connect regional mathematical models into a global net so that understanding the  unpredictability becomes possible. We make use of ``toy" perturbations due to the lack of preexisting ones, which should be found through experimental investigations. What we propose in this section is a small step in the mathematical approach to the complexity of the weather.


It is worth noting that the chaotification principles proposed in this paper are not specific for the Lorenz system (see Appendix), and they can be applied to other meteorological models as well, without any restrictions on the dimension and the number of the coupled systems. For example, one can consider the Lorenz model of general circulation of the atmosphere \cite{Lorenz84}
\begin{eqnarray}
\begin{array}{l} \label{Lorenz_general}
\displaystyle \frac{dX_1}{dt} = - X_2^2 -X_3^2 -\tilde{a}X_1+\tilde{a}F \\
\displaystyle \frac{dX_2}{dt}= X_1 X_2-\tilde{b}X_1X_3-X_2+G \\
\displaystyle \frac{dX_3}{dt} = \tilde{b} X_1 X_2 +X_1 X_3 - X_3,
\end{array}
\end{eqnarray}
where $X_1$ represents the strength of a large scale westerly wind current, $X_2$ and $X_3$ represent the cosine and sine phases of a chain of superposed large-scale eddies, the parameter $F$ represents the external-heating contrast, and $G$ represents the heating contrast between oceans and continents. The coefficient $\tilde{b},$ if greater than unity, allows the displacement to occur more rapidly than the amplification, and the coefficient $\tilde{a},$ if less than unity, allows the westerly current to damp less rapidly than the eddies \cite{Broer02,Lorenz84,Masoller95,Roebber95}. 

For $\tilde a >0$ and $\tilde b>-1,$ let us take into account the Lyapunov function $V(X)=X_1^2+X_2^2+X_3^2,$ and set  $\overline{a} = \min \left\{ \tilde{a},1 \right\},$ $\overline b= \sqrt{\tilde{a}^2F^2+G^2}.$  One can verify that
$ V'_{(\ref{Lorenz_general})}(X)  \le -2 \overline{a} \left(X_1^2+X_2^2+X_3^2\right) + 2 \overline{b} \sqrt{X_1^2+X_2^2+X_3^2} $ and $\displaystyle \left\|\frac{\partial V}{\partial X}\right\|= 2\sqrt{X_1^2+X_2^2+X_3^2}.$  Therefore, the conditions of Theorem \ref{butterfly_boundedness1a}, which is mentioned in the Appendix, are satisfied with $a(r)=r^2,$ $b(r)=2\overline{a}r^2-2\overline{b}r,$ $c(r)=2r,$ $M_0=1$ and $B=\displaystyle \frac{1+\overline b}{\overline a}.$ Consequently, our theory is also applicable to the Lorenz model of general circulation of the atmosphere.


\section{Conclusion}

In the present study, we investigate the dynamics of unidirectionally coupled Lorenz systems. It is rigorously proved that chaos can be extended from one Lorenz system to another. The extension of period-doubling cascade and sensitivity, which is the main ingredient of chaos, are shown both theoretically and numerically. Besides, the emergence of cyclic chaos, intermittency, and the concepts of self-organization and synergetics are considered for interconnected Lorenz systems. The results are valid if the drive Lorenz system is chaotic and the response system is non-chaotic, but admits a global asymptotically stable equilibrium or a globally attracting limit cycle. Our approach can give a light on the question about how global weather processes have to be described through mathematical models. A possible connection of the presented results with the global weather unpredictability is provided in the paper. The usage of our approach for the investigation of global weather unpredictability is a one more small step in the mathematical approach to the complexity of the weather. This is not a modelling of the atmosphere, but rather an effort to explain how the weather unpredictability can be arranged over the Earth on the basis of the Lorenz's meteorological model. In fact, this is also true for other meteorological models, since mathematical properties of stability, attraction and chaotic attractors are common for all models. It is shown that our results can be used for the Lorenz model of general circulation of the atmosphere \cite{Lorenz84} too. The question whether the overlapping of two chaotic dynamics may produce regularity can also be considered in future investigations. We guess that it is not possible, but an analysis has to be made.

\section*{Acknowledgments}

The authors wish to express their sincere gratitude to the referees for the helpful criticism and valuable suggestions, which helped to improve the paper significantly.

The second author is supported by the 2219 scholarship programme of T\"{U}B\.{I}TAK, the Scientific and Technological Research Council of Turkey.


\section{Appendix: The Mathematical Background}  \label{mathematical_background}

In our theoretical discussions, we consider more general coupled systems, which are not necessarily Lorenz systems. We will denote by $\mathbb R$ and $\mathbb N$ the sets of real numbers and natural numbers, respectively, and we will make use of the usual Euclidean norm for vectors.

Let us consider the autonomous systems
\begin{eqnarray} \label{system1_lorenz}
\frac{dx}{dt}=F(x),
\end{eqnarray}
and
\begin{eqnarray} \label{system3}
\frac{du}{dt}=f(u),
\end{eqnarray}
where $t\ge 0$ and the functions $F:\mathbb R^m \to \mathbb R^m$ and $f:\mathbb R^n \to \mathbb R^n$ are continuous in their arguments.

We perturb system  (\ref{system3}) with the solutions of (\ref{system1_lorenz}) and obtain the system in the form,
\begin{eqnarray} \label{system2_lorenz}
\frac{dy}{dt}=f(y)+\mu g(x(t)),
\end{eqnarray}
where the real number $\mu$ is nonzero and the function $g:\mathbb R^m \to \mathbb R^n$ is continuous. It is worth noting that the systems (\ref{lorenz_system}), (\ref{nonperturbed_lorenz_system}) and (\ref{perturbed_lorenz_system}) are in the form of (\ref{system1_lorenz}), (\ref{system3}) and (\ref{system2_lorenz}), respectively. 

We mainly assume that system (\ref{system1_lorenz}) possesses a chaotic attractor, let us say a set in $\mathbb R^m.$ Fix $x_0$ from the attractor and take a solution $x(t)$ of (\ref{system1_lorenz}) with $x(0)=x_0.$ Since we use the solution $x(t)$ as a perturbation in (\ref{system2_lorenz}), we call it as \textit{chaotic function}. Chaotic functions may be irregular as well as regular (periodic and unstable) \cite{Feigenbaum80,Lorenz63,Sander11,Sander12,Sch05,Wiggins88}. 
 
Our purpose is the prove rigorously the extension of chaos from system (\ref{system1_lorenz}) to system (\ref{system2_lorenz}). In our theoretical discussions, we request the existence of a bounded positively invariant region for system (\ref{system2_lorenz}). Such an invariant region can be achieved by different methods and one of them is mentioned in the next part. We will show the extension of sensitivity and the existence infinitely many unstable periodic solutions in Subsections \ref{butterfly_sensitivity} and \ref{period-doubling_subsection}, respectively.
  
In the following parts, for a given solution $x(t)$ of system $(\ref{system1_lorenz}),$ we will denote by $\phi_{x(t)}(t,t_0,y_0)$  the unique solution of system (\ref{system2_lorenz}) satisfying the initial condition $\phi_{x(t)}(t_0,t_0,y_0)=y_0.$

\subsection{\textbf{Existence of a bounded positively invariant region}}
  
Making benefit of Lyapunov functions and uniform ultimate boundedness \cite{Rouche77,Yoshizawa75}, we present a method in Theorem \ref{butterfly_boundedness1a} for the existence of a bounded positively invariant set of system (\ref{system2_lorenz}). Then, we will apply this technique to the Lorenz system.

Solutions of system (\ref{system2_lorenz}) are uniformly ultimately bounded if there exists a number $B_0>0$ and corresponding to any number $\alpha>0$ there exists a number $T(\alpha)>0$  such that  $\left\|y_0\right\| \leq \alpha$ implies that for each solution $x(t)$ of system (\ref{system1_lorenz}) and $t_0\ge 0$ we have $\left\|\phi_{x(t)}(t,t_0,y_0)\right\|<B_0$ for all $t \ge t_0+T(\alpha).$

The following condition is required:
\begin{enumerate}
\item[\bf (A1)] There exists a positive number $M_g$ such that $\displaystyle \sup_{x\in\mathbb R^m} \left\|g(x)\right\| \leq M_g.$
\end{enumerate}

\begin{theorem} \label{butterfly_boundedness1a}
Suppose that condition $(A1)$ is fulfilled and there exists a Lyapunov function $V(x)$ defined on $\mathbb R^n$  such that $V(x)$ has continuous first order partial derivatives. Additionally, assume that there exists a number $B\ge 0$ such that the following conditions are satisfied on the region $\left\|x\right\|\ge B:$
\begin{enumerate}
\item[\bf (i)]  $V\left(x\right) \ge  a\left(\left\|x\right\|\right),$  where  $a(r)$ is a continuous, increasing function defined for $r \ge B$ which satisfies $a(B)>0$ and $a(r) \to \infty$ as $r \to \infty;$
\item[\bf (ii)] $V'_{(\ref{system3})}(x)\le -b\left(\left\|x\right\|\right),$  where $b(r)$ is an increasing function defined for $r \ge B$ which satisfies $b(B)>0;$  
\item[\bf (iii)] $\displaystyle \left\|\frac{\partial V}{\partial x}(x)\right\|\leq c\left(\left\|x\right\|\right),$ where $c(r)$ is a function defined for $r \ge B$ and there exists a positive number $M_0$ such that $0<c(r)\le M_0 b(r)$ for all $r\ge B.$ 
\end{enumerate}
Then,  for sufficiently small $\left|\mu\right|,$ the solutions of system $(\ref{system2_lorenz})$ are uniformly ultimately bounded.
\end{theorem}

\noindent \textbf{Proof.} Fix arbitrary numbers $t_0 \ge 0,$ $\alpha > 0$ and a solution $x(t)$ of system (\ref{system1_lorenz}). Take a number $\beta$ satisfying $0<\beta<b(B).$ We consider system $(\ref{system2_lorenz})$ with a nonzero number $\mu$ which satisfies the inequality $$\displaystyle \left|\mu\right|\leq \frac{1}{M_0M_g}\left(1-\frac{\beta}{b(B)}\right).$$

Our aim is to show the existence of  numbers $B_0>B$ and $T(\alpha)\ge 0,$ independent of $t_0,$ such that if $\left\|y_0\right\| \le \alpha,$ then $\left\|\phi_{x(t)}(t,t_0,y_0)\right\|<B_0$ for all $t \ge t_0+T(\alpha).$  

Consider an arbitrary $y_0\in \mathbb R^n$ such that $\left\|y_0\right\| \le \alpha.$ For the sake of brevity, let us denote $y(t)=\phi_{x(t)}(t,t_0,y_0).$ 
In the proof, both of the possibilities $\left\|y_0\right\|<B$ and $\left\|y_0\right\|\ge B$ will be considered. We start with the former. 

Let $M_V=\displaystyle \max_{\left\|x\right\|=B}V(x).$ Since $a(r) \to \infty$ as $r \to \infty,$  there exists a number $B_0> B$ such that $a(B_0) \ge M_V.$

Now, suppose that there exists a moment $s_1>t_0$ such that $\left\|y(s_1)\right\|\ge B_0.$ It is possible to find a moment $s_2$ satisfying $t_0<s_2<s_1$ such that $\left\|y(s_2)\right\|= B$ and $\left\|y(t)\right\|\ge B$ for all $t \in [s_2,s_1].$

Assumptions $(ii)$ and $(iii)$ imply for $s_2\le t \le s_1$ that
\begin{eqnarray*}
&&\displaystyle \frac{dV(y(t))}{dt}  = \displaystyle \frac{\partial V}{\partial x}(y(t)) \cdot \left(f(y(t))+\mu g(x(t)) \right) \\
&& \le -b\left(\left\|y(t)\right\|\right)  + \left|\mu\right|M_g c\left(\left\|y(t)\right\|\right) \\
&& \le (\left|\mu\right|M_0M_g-1)b(B) \\
&& \le -\beta,
\end{eqnarray*}
where  $``\cdot"$ denotes the scalar product.

The last inequality  implies that $V(y(s_1))< V(y(s_2)).$ On the other hand, by the help of assumption $(i),$ we have $V(y(s_2)) \le M_V \le a(B_0) \le V(y(s_1)).$ This is  a contradiction. Therefore, for all $t \ge t_0$ the inequality $\left\|y(t)\right\| < B_0$ is valid.

Next, we consider the possibility $\left\|y_0\right\|\ge B.$ 
Since the function $V(x)$ is continuous and $\left\|y_0\right\| \le \alpha,$ one can find a number $K(\alpha)>0$ such that  $V(y_0) \le K(\alpha).$ By means of condition $(i)$ used together with the inequality $\left\|y_0\right\|\ge B,$ we have that $K(\alpha)\ge a(B).$

Assume that there exists a moment $\overline{t}> t_0+\displaystyle \frac{K(\alpha)-a(B)}{\beta}$ such that $\left\|y(\overline{t})\right\| \ge B.$

If there exists  $t_1 \in [t_0,\overline{t}]$ such that $\left\|y(t_1)\right\| < B,$ then by means of uniqueness of solutions, using a similar discussion to the case $\left\|y_0\right\|<B$ considered above, one can show that for all $t\ge t_1$ the inequality $\left\|y(t)\right\|<B_0$ holds. On the other hand, if for all $t\in[t_0,\overline{t}]$ the inequality $\left\|y(t)\right\| \ge B$ is valid, then one can verify that the inequality 
\begin{eqnarray*}
V(y(\overline{t})) \leq  V(y_0)-\beta(\overline{t}-t_0) 
\end{eqnarray*}
holds.
Under the circumstances we attain that
\begin{eqnarray*}
&& a(B) \leq V(y_0)-\beta(\overline{t}-t_0)\leq K(\alpha) - \beta (\overline{t}-t_0) < a(B).
\end{eqnarray*}
This is a contradiction. Hence, for all $t>t_0+T(\alpha),$ where $T(\alpha)=\displaystyle \frac{K(\alpha)-a(B)}{\beta},$ we have $\left\|y(t)\right\|<B_0.$ Consequently, the solutions of system $(\ref{system2_lorenz})$ are uniformly ultimately bounded.  $\square$


Next, we shall verify the conditions of Theorem $\ref{butterfly_boundedness1a}$ for the Lorenz model. Let us consider the system (\ref{nonperturbed_lorenz_system}) with the parameters $\overline{\sigma}>0,$ $0<\overline r<\sqrt{2}-1,$ $\overline b>0,$ and take into account the Lyapunov function
 \[ V(u)=\displaystyle \frac{1}{\overline{\sigma}} u_1^2 + u_2^2 + u_3^2,\] where $u=(u_1,u_2,u_3) \in \mathbb R^3.$ 


Set $\gamma_1=\displaystyle \min \left\{1, \frac{1}{\overline{\sigma}}\right\}$ and define the function $a(r)$ through the formula $a(r)=\gamma_1 r^2.$ In that case, the   relation
$
V(u) \ge \gamma_1 (u^2_1+u^2_2+u^2_3) = a \left( \left\|u\right\|  \right)
$
holds. On the other hand, one can verify that
\begin{eqnarray*}
&& \displaystyle   V'_{(\ref{nonperturbed_lorenz_system})}(u) = \displaystyle \frac{2}{\overline{\sigma}} u_1 u_1' + 2u_2u_2' + 2u_3u_3' \\
&& =\displaystyle \frac{2}{\overline{\sigma}} u_1 [\overline{\sigma} (-u_1+u_2)] + 2u_2(-u_1u_3+\overline{r} u_1-u_2) + 2u_3(u_1u_2-\overline{b}u_3) \\
&& = 2(\overline{r}+1)u_1u_2-2u_1^2-2u_2^2-2\overline{b}u_3^2.
\end{eqnarray*}

Now, let $\gamma_2=\displaystyle \min \left\{1, 2-(\overline{r}+1)^2, 2\overline{b}\right\}.$ Making use of the identity 
\begin{eqnarray*}
2(\overline{r}+1)u_1u_2=\displaystyle u_1^2  +\displaystyle (\overline{r}+1)^2 u_2^2-\left[u_1-(\overline{r}+1) u_2\right]^2
\end{eqnarray*}
we attain the inequality
\begin{eqnarray*}
&& \displaystyle  V'_{(\ref{nonperturbed_lorenz_system})}(u)= -  \left[ u_1-(\overline{r}+1)u_2 \right]^2 - u_1^2 - \left[2-(\overline{r}+1)^2\right] u_2^2-2\overline{b}u_3^2 \\
&& \le  - u_1^2 - \left[2-(\overline{r}+1)^2\right] u_2^2-2\overline{b}u_3^2 \\
&& \le  - b\left(\left\|u\right\|\right),
\end{eqnarray*}
where the function $b(r)$ is defined through the formula $b(r)=\gamma_2 r^2.$ The last inequality validates the condition $(ii)$ of Theorem \ref{butterfly_boundedness1a}.

Furthermore, one can obtain that
\begin{eqnarray*}
 \left\|\frac{\partial V}{\partial u}(u)\right\| = 2 \sqrt{\frac{1}{\overline{\sigma}^2}u_1^2+u_2^2+u_3^2}  \leq c\left(\left\|u\right\|\right),
\end{eqnarray*}
where $\displaystyle c(r)=2\gamma_3 r$ and $\displaystyle \gamma_3=\max\left\{1,\frac{1}{\overline{\sigma}}\right\}.$ If we take $M_0=\displaystyle \frac{2\gamma_3}{\gamma_2},$ then the inequality $c(r)\le M_0 b(r)$ holds for all $r\ge 1.$ Consequently, for  $B=1,$ the conditions of Theorem \ref{butterfly_boundedness1a} are satisfied for system (\ref{nonperturbed_lorenz_system}) with the coefficients $\overline{\sigma}>0,$ $0<\overline r<\sqrt{2}-1$ and $\overline b>0.$

In the next section, we will continue with the extension of sensitivity, which can be considered as the unique ingredient of chaos for a set of bounded solutions \cite{Lorenz63,Rob95,Wiggins88}.  


\subsection{\textbf{Unpredictability analysis}} \label{butterfly_sensitivity}

Extension of the sensitivity feature through system  $(\ref{system2_lorenz})$ will be handled in the present part.  We shall begin with the meaning of the aforementioned property for systems $(\ref{system1_lorenz})$ and $(\ref{system2_lorenz}).$ The main result will be stated in Theorem \ref{sensitivity_thm_lorenz}.

System (\ref{system1_lorenz}) is called sensitive if there exist positive numbers $\epsilon_0$ and $\Delta$ such that for an arbitrary positive number $\delta_0$ and for each chaotic solution $x(t)$ of system $(\ref{system1_lorenz}),$ there exist a chaotic solution $\overline{x}(t)$ of the same system and an interval $J  \subset [0,\infty),$ with a length no less than $\Delta,$ such that $\left\|x(0)-\overline{x}(0)\right\|<\delta_0$ and $\left\|x(t)-\overline{x}(t)\right\| > \epsilon_0$ for all $t \in J.$  

Our main assumption is the existence of a bounded positively invariant set $\mathscr{K}$ for system $(\ref{system2_lorenz}).$ The existence of such an invariant set can be shown, for example, by using Theorem $\ref{butterfly_boundedness1a}.$

We say that system (\ref{system2_lorenz}) is sensitive if there exist positive numbers $\epsilon_1$ and $\overline{\Delta}$ such that for an arbitrary positive number $\delta_1,$ each $y_0\in \mathscr{K}$ and a chaotic solution $x(t)$ of (\ref{system1_lorenz}), there exist $y_1\in \mathscr{K},$ a chaotic solution $\overline{x}(t)$ of (\ref{system1_lorenz}) and an interval $J^1 \subset [0,\infty),$ with a length no less than $\overline{\Delta},$ such that $\left\|y_0-y_1\right\|<\delta_1$ and $\left\|\phi_{x(t)}(t,0,y_0)-\phi_{\overline{x}(t)}(t,0,y_1)\right\| > \epsilon_1$ for all $t \in J^1.$

The following assumptions are needed:
\begin{enumerate}
\item[\bf (A2)]  There exists a positive number $M_F$  such that 
$\displaystyle \sup_{x\in\mathbb R^m} \left\|F(x)\right\| \leq M_F;$ 
\item[\bf (A3)] There exists a positive number $L_f$ such that 
$\left\|f(y_1)-f(y_2)\right\| \leq L_f\left\|y_1-y_2\right\|$ for all $y_1,y_2\in \mathbb R^n;$
\item[\bf (A4)] There exists a positive number $L_g$ such that 
$\left\|g(x_1)-g(x_2)\right\| \ge L_g \left\|x_1-x_2\right\|$ for all $x_1,x_2 \in \mathbb R^m.$
\end{enumerate}

In the next theorem, the extension of sensitivity from system $(\ref{system1_lorenz})$ to system $(\ref{system2_lorenz})$ is considered.

\begin{theorem} \label{sensitivity_thm_lorenz}
Suppose that conditions $(A1)-(A4)$ hold. If system $(\ref{system1_lorenz})$ is sensitive, then the same is true for system $(\ref{system2_lorenz}).$
\end{theorem}

\noindent \textbf{Proof.} 
Fix an arbitrary positive number $\delta_1,$ $y_0\in \mathscr{K}$ and a chaotic solution $x(t)$ of (\ref{system1_lorenz}). Since system (\ref{system1_lorenz}) is sensitive, one can find $\epsilon_0>0$ and $\Delta>0$ such that for arbitrary $\delta_0>0$ both of the inequalities $\left\|x(0)-\overline{x}(0)\right\|<\delta_0$ and $\left\|x(t)-\overline{x}(t)\right\| > \epsilon_0,$ $t \in J,$ hold for some chaotic solution $\overline{x}(t)$ of $(\ref{system1_lorenz})$ and for some interval $J \subset [0,\infty),$ whose length is not less than $\Delta.$ 

Take an arbitrary $y_1\in \mathscr{K}$ such that $\left\|y_0-y_1\right\|<\delta_1.$ For the sake of brevity, let us denote $y(t)=\phi_{x(t)}(t,0,y_0)$ and   $\overline{y}(t)=\phi_{\overline{x}(t)}(t,0,y_1).$ 

It is worth noting that there exist positive numbers $K_0$ and $H_0$ such that $\left\|y(t)\right\|, \left\|\overline y(t)\right\| \le K_0$ for all $t\ge 0$ and  $\displaystyle \sup_{t \ge 0} \left\|x(t)\right\| \leq H_0$ for each chaotic solution $x(t)$ of system $(\ref{system1_lorenz}).$
 
Our aim is to determine positive numbers $\epsilon_1,$ $\overline{\Delta}$ and an interval $J^1\subset [0,\infty)$ with length $\overline{\Delta}$ such that the inequality $\left\|y(t)-\overline{y}(t)\right\| > \epsilon_1$ holds for all $t \in J^1.$

Since the derivative of each chaotic solution $x(t)$ of (\ref{system1_lorenz}) lies inside the tube with radius $M_F,$ the collection of chaotic solutions of system $(\ref{system1_lorenz})$ is an equicontinuous family on $[0,\infty).$ Suppose that
$
 g(x) = \left(  g_1(x),  g_2(x), \ldots,  g_n(x) \right),
$
where each $g_j,$ $1 \leq j \leq n,$ is a real valued function. Making use of the uniform continuity of the function $\overline{g}: \mathbb R^m \times \mathbb R^m  \to \mathbb R^n,$ defined as $\overline{g}(\nu_1,\nu_2)=g(\nu_1)-g(\nu_2),$ on the compact region 
$
\mathscr{R}=\left\{(\nu_1,\nu_2) \in \mathbb R^m \times \mathbb R^m : \left\|\nu_1\right\| \leq H_0, \left\|\nu_2\right\| \leq H_0 \right\} 
$
together with the equicontinuity of the collection of chaotic solutions of $(\ref{system1_lorenz}),$ one can verify that the collection $\mathscr{F}$ consisting of the functions of the form
$g_j(x_1(t))-g_j(x_2(t)), 1\leq j\leq n,$ where $x_1(t)$ and $x_2(t)$ are chaotic solutions of system $(\ref{system1_lorenz}),$ is an equicontinuous family on $[0,\infty).$

According to the equicontinuity of the family $\mathscr{F},$ one can find a positive number $\tau<\Delta,$ which is independent of $x(t)$ and $\overline{x}(t),$  such that for any $t_1,t_2\in [0,\infty)$ with $\left|t_1-t_2\right|<\tau,$ the inequality 
\begin{eqnarray} \label{sensitivity_proof_1}
\begin{array}{l}
\left| \left(g_j\left(x(t_1)\right) - g_j\left(\overline{x}(t_1)\right)  \right) - \left(g_j\left(x(t_2)\right) - g_j\left(\overline{x}(t_2)\right)  \right)   \right| <\displaystyle \frac{L_g\epsilon_0}{2n}
\end{array}
\end{eqnarray}
holds for all $1\leq j \leq n.$

Condition $(A4)$ implies that $\left\|g(x(t))-g(\overline{x}(t)) \right\| \geq L_g \left\|x(t)-\overline{x}(t)\right\|,$ $t \in J.$  Therefore, for each $t\in J,$ there exists an integer $j_0,$ $1 \leq j_0 \leq n,$ which possibly depends on $t,$ such that 
\begin{eqnarray}
\begin{array}{l}
\left|g_{j_0}(x(t))-g_{j_0}(\overline{x}(t))\right|   \geq \displaystyle \frac{L_g}{n} \left\|x(t)-\overline{x}(t)\right\| \nonumber.
\end{array}
\end{eqnarray}
Otherwise, if there exists $s\in J$ such that for all $1\leq j\leq n$ the inequality  
\begin{eqnarray}
\begin{array}{l}
\left|g_{j} \left(x\left(s\right)  \right)-g_{j}(\overline{x}(s)) \right|< \displaystyle \frac{L_g}{n} \left\|x(s)-\overline{x}(s)\right\| \nonumber
\end{array}
\end{eqnarray}
holds, then one encounters with a contradiction since
\begin{eqnarray*}
\left\|g(x(s))-g(\overline{x}(s))  \right\|   \leq \sum_{j=1}^{n}\left| g_{j}(x(s))-g_{j}(\overline{x}(s)) \right|  < L_g \left\|x(s)-\overline{x}(s)\right\|. 
\end{eqnarray*}

Denote by $s_0$ the midpoint of the interval $J,$ and let $\displaystyle \theta=s_0-\tau/2.$ There exists an integer $j_0,$  $1 \leq j_0 \leq n,$ such that 
\begin{eqnarray}
\begin{array}{l} \label{sensitivity_proof_2}
\left|g_{j_0}(x(s_0))-g_{j_0}(\overline{x}(s_0))\right|   \geq \displaystyle\frac{L_g}{n} \left\|x(s_0)-\overline{x}(s_0)\right\| > \displaystyle\frac{L_g\epsilon_0}{n}. 
\end{array}
\end{eqnarray}
On the other hand, making use of the inequality (\ref{sensitivity_proof_1}) it can be verified for all $t \in \left[\theta, \theta+\tau\right]$ that
\begin{eqnarray*}
&& \left|g_{j_0}\left(x(s_0)\right) - g_{j_0}\left(\overline{x}(s_0)\right) \right| - \left|g_{j_0}\left(x(t)\right) - g_{j_0}\left(\overline{x}(t)\right) \right| \\
&& \leq \left| \left(g_{j_0}\left(x(t)\right) - g_{j_0}\left(\overline{x}(t)\right)  \right) - \left(g_{j_0}\left(x(s_0)\right) - g_{j_0}\left(\overline{x}(s_0)\right)  \right)   \right| \\
&&<\frac{L_g\epsilon_0}{2n}.
\end{eqnarray*}
Therefore, by means of $(\ref{sensitivity_proof_2}),$ we have that the inequality
\begin{eqnarray} \label{sensitivity_proof_3}
\begin{array}{l} 
 \left|g_{j_0}\left(x(t)\right) - g_{j_0}\left(\overline{x}(t)\right) \right| > \left|g_{j_0}\left(x(s_0)\right) - g_{j_0}\left(\overline{x}(s_0)\right) \right|  - \displaystyle \frac{L_g\epsilon_0}{2n} > \displaystyle \frac{L_g\epsilon_0}{2n}
 \end{array}
\end{eqnarray}
is valid for  $t\in \left[\theta, \theta+\tau\right].$

One can find numbers $s_1, s_2, \ldots, s_n \in [\theta,\theta+\tau]$ such that
\begin{eqnarray*}
\displaystyle\int^{\theta + \tau}_{\theta} \left[g(x(s))-g(\overline{x}(s))\right] ds   = \Big(  
\tau \left[g_1(x(s_1))-g_1(\overline{x}(s_1))\right],  
\tau \left[g_2(x(s_2))-g_2(\overline{x}(s_2))\right],  \\ 
\ldots,  
\tau \left[g_n(x(s_n))-g_n(\overline{x}(s_n))\right] 
  \Big).
\end{eqnarray*}
By using the inequality $(\ref{sensitivity_proof_3}),$ we attain that
\begin{eqnarray*}  
\left\|\displaystyle\int^{\theta + \tau}_{\theta} \left[g(x(s))-g(\overline{x}(s))\right] ds \right\| \geq \tau  \left|g_{j_0}(x(s_{j_0}))-g_{j_0}(\overline{x}(s_{j_0}))\right| > \displaystyle \frac{\tau  L_g \epsilon_0}{2n}.
\end{eqnarray*}
The relation
\begin{eqnarray*}
y(t)-\overline{y}(t)  = (y(\theta)-\overline{y}(\theta)) + \displaystyle\int^{t}_{\theta}  \left[ f(y(s))-f(\overline{y}(s)) \right]   ds  + \displaystyle\int^{t}_{\theta} \mu [g(x(s))-g(\overline{x}(s))]  ds,   ~t\in [\theta,\theta+\tau]
\end{eqnarray*}
yields
\begin{eqnarray*} 
&& \left\| y(\theta+\tau)-\overline{y}(\theta+\tau) \right\| \geq  \left|\mu\right| \left\|\displaystyle\int^{\theta+\tau}_{\theta}  [g(x(s))-g(\overline{x}(s))] ds \right\| \\
&& - \left\| y(\theta)-\overline{y}(\theta) \right\| - \displaystyle\int^{\theta+\tau}_{\theta}  L_f\left\| y(s)-\overline{y}(s) \right\| ds \\
&& > \displaystyle \frac{ \left| \mu \right|  \tau  L_g \epsilon_0}{2n} - \left\| y(\theta)-\overline{y}(\theta) \right\| - \displaystyle\int^{\theta+\tau}_{\theta}  L_f\left\| y(s)-\overline{y}(s) \right\| ds.
\end{eqnarray*}
The last inequality implies that
\begin{eqnarray*}
&& \displaystyle \max_{t\in [\theta,\theta+\tau]}\left\| y(t)-\overline{y}(t)\right\|  \geq \left\| y(\theta+\tau)-\overline{y}(\theta+\tau) \right\| \\
&& > \frac{\left|\mu\right|\tau L_g \epsilon_0}{2n}  - (1+ \tau L_f) \displaystyle \max_{t\in [\theta,\theta+\tau]}\left\| y(t)-\overline{y}(t) \right\|.
\end{eqnarray*}
Therefore,  
$
\displaystyle \max_{t\in [\theta,\theta+\tau]}\left\| y(t)-\overline{y}(t) \right\| > \frac{\left|\mu\right|\tau L_g \epsilon_0}{2n(2+\tau L_f)}.
$

Suppose that  
$
\displaystyle \max_{t \in [\theta,\theta+\tau]} \left\|y(t)-\overline{y}(t)\right\|  = \left\|y(\xi)-\overline{y}(\xi)\right\| 
$
for some $\xi \in [\theta, \theta+\tau].$ 
Define
\begin{displaymath}
\overline{\Delta}=\min \displaystyle \left\{ \frac{\tau}{2}, \frac{\left|\mu\right|\tau L_g \epsilon_0}{8n(K_0L_f+M_g\left|\mu\right|)(2+\tau L_f)}   \right\}
\end{displaymath}
and let
\begin{displaymath}
\theta^1=\left\{\begin{array}{ll} \xi, & ~\textrm{if}~  \xi \leq \theta + \tau/2   \\
\xi - \overline{\Delta}, & ~\textrm{if}~  \xi > \theta + \tau/2  \\
\end{array} \right. .\nonumber
\end{displaymath}

For $t\in [\theta^1, \theta^1+\overline{\Delta}],$  by favour of the  equation
\begin{eqnarray*}
y(t)-\overline{y}(t) = (y(\xi)-\overline{y}(\xi)) + \displaystyle\int^{t}_{\xi} \left[f(y(s))-f(\overline{y}(s))\right] ds  + \displaystyle\int^{t}_{\xi} \mu [g(x(s))-g(\overline{x}(s))]  ds,
\end{eqnarray*}
one can obtain that
\begin{eqnarray*} 
&& \left\|y(t)-\overline{y}(t)\right\|  \geq  \left\|y(\xi)-\overline{y}(\xi)\right\|  - \left|  \displaystyle\int^{t}_{\xi} L_f \left\| y(s)-\overline{y}(s)\right\| ds  \right|\\ 
&& - \left|\mu\right| \left|  \displaystyle\int^{t}_{\xi} \left\|  g(x(s))-g(\overline{x}(s))  \right\| ds  \right| \\
&& > \displaystyle\frac{\left|\mu\right|\tau L_g \epsilon_0}{2n(2+\tau L_f)} -2\overline{\Delta} \left(K_0L_f+M_g\left|\mu\right|\right) \\
&& \geq \displaystyle\frac{\left|\mu\right|\tau L_g \epsilon_0}{4n(2+\tau L_f)}.
\end{eqnarray*}

The length of the interval $J^1=[\theta^1, \theta^1+\overline{\Delta}]$ does not depend on $x(t),$ $\overline{x}(t),$ and for $t \in J^1$ the inequality
$
\left\|y(t)-\overline{y}(t)\right\| > \epsilon_1
$ 
holds, where $\epsilon_1=\displaystyle \frac{\left|\mu\right|\tau L_g \epsilon_0}{4n(2+\tau L_f)}.$ Consequently, system $(\ref{system2_lorenz})$ is sensitive.  $\square$

\subsection{\textbf{Existence of unstable periodic motions}} \label{period-doubling_subsection}

Assume  that system (\ref{system1_lorenz}) admits a period-doubling cascade. That is, there exists an equation
\begin{eqnarray}
\begin{array}{l}
x'=G(x,\lambda),  \label{period_doubling1}
\end{array}
\end{eqnarray}
where $\lambda$ is a parameter and the function $G:\mathbb R^m \times \mathbb R \to \mathbb R^m$ is such that for some finite number $\lambda_{\infty},$   $G(x,\lambda_{\infty})$ is equal to the function $F(x)$ in the right hand side of system (\ref{system1_lorenz}).

System (\ref{system1_lorenz}) is said to admit a period-doubling cascade \cite{Feigenbaum80,Sander11,Sander12,Zelinka} if there exists a sequence of period-doubling bifurcation values $\left\{\lambda_j\right\}_{j\in\mathbb N}$ satisfying $\lambda_j \to \lambda_{\infty}$ as $j \to \infty$ such that as the parameter $\lambda$ increases or decreases through $\lambda_j$ system (\ref{period_doubling1}) undergoes a period-doubling bifurcation for each $j\in\mathbb N.$ As a consequence, at the parameter value $\lambda=\lambda_{\infty},$ there exist infinitely many unstable periodic solutions of system (\ref{period_doubling1}), and hence of system $(\ref{system1_lorenz}),$ all lying in a bounded region.

Now, let us introduce the following definition \cite{Yoshizawa75}. We say that the solutions of the non-autonomous system (\ref{system2_lorenz}), with a fixed $x(t),$  are ultimately bounded if there exists a number $B>0$ such that for every solution $y(t),$ $y(t_0)=y_0,$ of system $(\ref{system2_lorenz}),$ there exists a positive number $R$ such that the inequality $\left\| y(t) \right\|<B$ holds for all $t\geq t_0+R.$

We say that system (\ref{system2_lorenz}) replicates the period-doubling cascade of system (\ref{system1_lorenz}) if for each periodic solution $x(t)$ of (\ref{system1_lorenz}), system (\ref{system2_lorenz}) admits a periodic solution with the same period.  

The following condition is required in the next theorem, which can be verified by using Theorem $15.8$ \cite{Yoshizawa75}.

\begin{enumerate}
\item[\bf (A5)] Solutions of system (\ref{system2_lorenz}) are ultimately bounded by a bound common for all $x(t).$
\end{enumerate}

\begin{theorem}\label{period-doubling_theorem}
If conditions $(A1)-(A5)$ hold, then system (\ref{system2_lorenz}) replicates the period-doubling cascade of system (\ref{system1_lorenz}).
\end{theorem}

It is worth noting that the instability of the infinite number of periodic solutions of system $(\ref{system2_lorenz})$ is ensured by Theorem $\ref{sensitivity_thm_lorenz}.$


\end{document}